\documentclass[journal,10pt]{IEEEtran}
\usepackage{graphicx,amsfonts,amsmath, amssymb,array,url}
\usepackage{subfigure,setspace,multirow}
\usepackage{epstopdf}
\usepackage{hhline}
\usepackage{amsthm}
\usepackage{cite}
\usepackage{soul}
\usepackage[english]{babel}
\usepackage{atbegshi}
\AtBeginDocument{\AtBeginShipoutNext{\AtBeginShipoutDiscard}}
\usepackage{relsize}
\newtheorem{thm}{Theorem}

\begin{document}
\title{A Unified Framework for Multi-Hop Wireless Relaying with Hardware Impairments}
\author{Ehsan Soleimani-Nasab, and Sinem Coleri, \IEEEmembership{Fellow, IEEE}}
\thanks{Manuscript received May 13, 2023; revised July 18, 2023; accepted November 6, 2023.
The associate editor coordinating the review of this paper and approving it for publication was Prof. H. Lin.

E. Soleimani-Nasab is with the Department of Electrical and Computer Engineering, Graduate University of Advanced Technology, Kerman 7631885356, Iran (e-mail: ehsan.soleimani@kgut.ac.ir, enasab@ku.edu.tr).

 S. Coleri is with the Department of Electrical and Electronics Engineering, Ko\c{c} University, Istanbul 34450, Turkey (e-mail: scoleri@ku.edu.tr).

 This work was supported by the Scientific and Technological Research Council of Turkey (TUBITAK) under the research fellowship TUBITAK-BIDEB 2221.

 Digital Object Identifier XX/TVT.2023.XX

Copyright (c) 2023 IEEE. Personal use of this material is permitted. However, permission to use this material for any other purposes must be obtained from the IEEE by sending a request to pubs-permissions@ieee.org.
}

\maketitle
\makeatletter
\long\def\@makecaption#1#2{\ifx\@captype\@IEEEtablestring%
\footnotesize\begin{center}{\normalfont\footnotesize #1}\\
{\normalfont\footnotesize\scshape #2}\end{center}%
\@IEEEtablecaptionsepspace
\else
\@IEEEfigurecaptionsepspace
\setbox\@tempboxa\hbox{\normalfont\footnotesize {#1.}~~ #2}%
\ifdim \wd\@tempboxa >\hsize%
\setbox\@tempboxa\hbox{\normalfont\footnotesize {#1.}~~ }%
\parbox[t]{\hsize}{\normalfont\footnotesize \noindent\unhbox\@tempboxa#2}%
\else
\hbox to\hsize{\normalfont\footnotesize\hfil\box\@tempboxa\hfil}\fi\fi}
\makeatother
 \begin{abstract}
Relaying increases the coverage area and reliability of wireless communications systems by mitigating the fading effect on the received signal.
Most technical contributions in the context of these systems assume ideal hardware (ID) by neglecting the non-idealities of the transceivers, which include phase noise, in-phase/quadrature mismatch and high power amplifier nonlinearities. These non-idealities create distortion on the received signal by causing variations in the phase and attenuating the amplitude. The resulting deterioration of the performance of wireless communication systems is further magnified as the frequency of transmission increases. In this paper, we investigate the aggregate impact of hardware impairments (HI) on the general multi-hop relay system using amplify-and-forward (AF) and decode-and-forward (DF) relaying techniques over a general $\bf H$-fading model. $\bf H$-fading model includes free space optics, radio frequency, millimeter wave, Terahertz, and underwater fading models. Closed-form expressions of outage probability, bit error probability and ergodic capacity are derived in terms of $\bf H$-functions. Following an asymptotic analysis at high signal-to-noise ratio (SNR), practical optimization problems have been formulated with the objective of finding the optimal level of HI subject to the limitation on the total HI level. The analytical solution has been derived for the Nakagami-$m$ fading channel which is a special case of $\bf H$-fading for AF and DF relaying techniques. The overall instantaneous signal-to-noise-plus-distortion ratio has been demonstrated to reach a ceiling at high SNRs which has a reciprocal proportion to the HI level of all hops' transceivers on the contrary to the ID.
   \end{abstract}


\setcounter{page}{1}
\section{Introduction}
Fifth generation (5G) and sixth generation (6G) wireless communication systems play a valuable role in increasing the quality of human life by providing a very high data rate, very low latency and very high reliability. According to the recent Ericsson report, the global number of 5G subscriptions is expected to exceed 5 billion by the end of 2028 with over 300 exabyte (EB) data traffic on mobile subscriptions \cite{ericsson}. To achieve the aforementioned goals, various frequency bands have been regularized by Federal Communications Commission (FCC) including free space optics (FSO), radio frequency (RF), millimeter wave (MMW), and Terahertz (THz) technologies \cite{Chowdhury}.
In these frequency bands, propagation loss and fading are extremely high, leading to signal blockage due to the absorption of the signal energy and low transmission power at higher frequencies.
 multi-hop (MH) relaying provides a low cost solution to extend the transmission coverage in wireless links and diminish fading since the fading variance depends on the distance between transmitter and receiver. The two most widely used protocols in cooperative communications are amplify-and-forward (AF), and decode-and-forward (DF)
  where in the {former the relay sends the received signal after amplifying, while in the latter the relay first decodes the received signal and then sends it toward the destination} \cite{relay2022}. {In recent years, several application scenarios for MH wireless networks have been investigated including (i) cellular radio/optical networks to extend the coverage range using relaying; (ii) wireless mesh networks for providing broadband internet services without the need of expensive cable infrastructures, in particular in areas sparsely populated as well as in urban areas where the cost of laying cables is very high; (iii) multi-satellite systems for deep space communications; (iv) vehicular communications with mobile relay nodes, which is a special case of mobile ad-hoc networks; and (v) wireless sensor networks as part of internet of things, which offer large geographical areas with connectivity without having direct physical access to each sensor node \cite{Shen}.}

The first set of relaying based studies assume ideal hardware (ID), where the hardware impairments (HI) in the hardware of the RF and FSO link, including phase noise \cite{Costa-phase, Gappmair_HPA}, in-phase/quadrature (I/Q) mismatch \cite{Qi-IQ, Changle_IQ} and high power amplifier (HPA) nonlinearities \cite{Dardari, Zedini_HI}, are ignored.
 In \cite{jocn2022}, the performance of relay-assisted MH systems has been studied for the configuration where multiple RF and FSO links were cascaded using either AF or DF relaying with ID. \cite{MultihopF} derives the analytical expressions of outage probability (OP) and bit error probability (BEP) performance for the MH relaying assuming $\mathcal{F}$ distributed {atmospheric turbulence with pointing  errors (PE) along with high signal-to-noise ratio (SNR) analysis of the OP and BEP.} \cite{MHTHz} provides the closed-form expressions of the OP and BEP of MH THz wireless relaying systems. Also, the diversity order through an asymptotic analysis of the OP and BEP was obtained. \cite{EGG} studies the MH relaying of underwater fading over mixture Exponential-generalized Gamma (mEGG) distribution, where exact and asymptotical expressions of OP and BEP are presented for both type of detection techniques (i.e., heterodyne detection (HD) and direct detection (DD)). However, assuming ID by neglecting hardware impairment distortion can cause very deceitful results in the analysis of high-rate systems due to the destructive effects of HI on the instantaneous signal-to-noise-plus-distortion ratio (SNDR). Among the HI, when a signal is frequency multiplied, the phase noise increases around six dB for every doubling. I/Q mismatch rotates the phase and attenuates the amplitude of the desired constellation. Furthermore, an image interference is created by the mirrored subcarrier, which causes a bit-error-rate (BER) floor. In comparison to linear HPAs, HPA non-linearities increase the BER; for extreme non-linearities, an irreducible error floor appears.

The second set of relaying based studies incorporate HI into their analysis due to the wide deployment of inexpensive hardware and the usage of higher frequency bands.
\cite{Bjornson} considers a generalized system model to characterize the impact of HI in dual-hop relay networks over Nakagami-$m$ fading channels.
Closed-form expressions are derived for the OP and the ergodic capacity (EC) followed by high SNR analysis based on the derived SNDR. Some design guidelines for both AF and DF relaying systems based on the fundamental limits in the HI systems are also discussed. In the context of mixed RF and FSO systems, \cite{Balti} evaluates the impact of the HI on a dual-hop RF-FSO relay network. Assuming Rayleigh fading and Malaga turbulence with PE respectively for the RF and FSO links, the closed-form expressions for the OP and BEP along with high SNR analysis including the diversity gains are presented. \cite{Jiayi} studies dual-hop mixed THz-FSO relaying systems assuming the non-ID for AF relaying systems. The exact and asymptotic OP of the systems are derived over $\boldsymbol\alpha-\boldsymbol\mu$ fading and double generalized Gamma (DGG) turbulence with the PE along with the analytical expression of the diversity gain. {The effect of RF impairments, which is modeled as independent and identically distributed (i.i.d.) additive Gaussian noise, is investigated in \cite{Boulogeorgos,Hieu,Zhang1,Mouchtak,Solanki}.
In \cite{Boulogeorgos}, authors investigate the effects of I/Q imbalance at the Tx/Rx over $N\ast$ Nakagami-$m$ fading channels. Closed-form expressions of OP for single and multi-carrier communication systems have been derived in the terms of H-function. In \cite{Hieu}, a secrecy analysis of multi-hop hardware-impaired relaying system with DF relaying is given for Rayleigh fading channels.  Exact closed-form expressions and the asymptotic OP for three selection protocols are derived.
In \cite{Zhang1}, authors present an analysis of dual-hop CSI-assisted AF relaying communications assuming $\mathcal{F}$-distribution. The approximated OP, BEP, EC, and the effective capacity are further obtained in closed-form. \cite{Mouchtak} gives a performance analysis of I/Q mismatch with HI Over H-fading channel. Closed-form expressions and asymptotic expressions for the OP, BEP, and EC are obtained. \cite{Solanki} studies the performance of a cognitive radio network with multiple DF relays under the impact of HI. Exact closed-form expression for the OP employing selection cooperation over i.n.i.d. Rayleigh fading channels are derived.} In the context of MH relaying, \cite{Alraddady} proposes a MH full-duplex communication systems under hardware manufacturing defects and self interference over independent but non-identically distributed (i.n.i.d.) Nakagami-$m$ fading channels. The approximated OP, BEP, and EC based on the Euler numerical technique are derived. While these studies have improved our knowledge on the performance characterization of relaying systems with HI, they are limited in terms of the number of hops and fading channel models. \cite{Bjornson, Balti, Jiayi} consider the aggregate impact of HI for the dual-hop relaying without considering MH relaying. On the other hand, \cite{Alraddady} provides an analytical framework for the MH relaying considering Nakagami-$m$ fading only, which does not necessarily apply to the FSO, MMW and THz frequency bands.

In this paper, we propose a unified framework to the impact of aggregate HI on the performance of the MH relaying in 5G and 6G networks over high frequency bands by considering the generalized i.n.i.d. $\bf H$-fading model, for the first time in the literature. {Since the framework considers a general fading model which consists many fading models in RF, THz, FSO, and MMW systems, arbitrary receiver types, arbitrary number of links, and arbitrary level of impairment, we call the proposed model as unified.} {The aggregate HI considers the phase noise, I/Q mismatch and HPA nonlinearities considering soft envelope limiter, solid state power amplifier, traveling wave tube amplifier and ideal soft-limiter amplifier models altogether.} $\bf H$-fading model has been demonstrated to represent accurately many of the fading channels at high frequency bands, including RF, FSO, THz and MMW bands, and covering DGG {plus PE}, Fisher-Snedecor $\mathcal{F}$, extended generalized-K (EGK), $\boldsymbol\alpha-\boldsymbol\mu$ {plus PE}, Malaga {plus PE}, mEGG \cite{Jeong} and $N\ast$ Nakagami-$m$ channels \cite{KaragiannidisN}.
New closed-form expressions are derived for the OP and EC under aggregate HI.\footnote{
{Each metric has been used to quantify the performance gains/impairments and to comprehend how factors arising from design/implementation (e.g. transmitter noise, channel noise, receiver noise, diversity, multipath fading) affect overall system performance. In the context of relaying, these metrics were frequently used in the literature (Please see \cite{jocn2022}-\cite{KaragiannidisN} for more details).}}
The original contributions of this paper are given as follows:

\begin{itemize}

\item We propose a generalized framework for the inclusion of aggregate HI and $\bf H$-fading model in the performance analysis of the MH relaying
systems over high frequency bands, for the first time in the literature.

\item We derive the closed-form expressions of the instantaneous end-to-end (E2E) SNDR, E2E OP and EC for both AF and DF MH relaying, for the first time in the literature. These expressions are derived considering both the inductive argument method and Cauchy's residue theorem where the performance of a $n+1$-hop relaying system is characterized by that of an $n$-hop relaying system for $n=2,...,N$.

\item  We derive the dependence of the diversity gain of the AF relaying on the parameters of the $\bf H$-fading and number of links, for the first time in the literature. The diversity gain expression is derived at high SNRs based on an asymptotic OP expression employing both inductive argument method and Cauchy's residue theorem.

\item Using the proposed framework, we formulate the optimization problem with the objective function of minimizing the OP, decision variables of the HI levels of the nodes and constraint on the total HI levels for AF and DF relaying techniques, for the first time in the literature. The cost of the hardware determines the level of hardware impairment such that the lower the cost, the higher the HI. We obtain closed-form analytical solutions of the optimal level of HI for Nakagami-$m$ fading channels, which is a special case of $\bf H$-fading.
\end{itemize}

We organize the rest of the paper as follows. Section II describes the system model. Section III and IV provide the OP, BEP, and EC performance of the MH system for AF and DF relaying protocols, respectively. Section V describes the formulation of the optimization problem for determining the optimal level of HI to minimize the OP. Section VI presents some numerical results. Finally, Section VII concludes the paper.

\textit{Notation}: Throughout this paper, the operator ${Pr}(.)$ denotes probability, while $\mathbb{E}_{\{.\}}\{.\}$  stands for expectation operator. $\Gamma(n)=\textstyle \int_{0}^{\infty} e^{-t} t^{n-1}\, dt$ is the Gamma function \cite[Eq. (8.310.1)]{Gradshteyn:2007}, and $\overline {{\Gamma}} $ is the average SNR. $\mathcal{CN}(\mu,\sigma^2)$ shows a complex Gaussian random variable (RV) with mean of $\mu$ and variance of $\sigma^2$, and the operator $\sim$ means distributed as.
\section{System Model}
\begin{figure*}
\centering
\includegraphics[keepaspectratio,width=15cm]{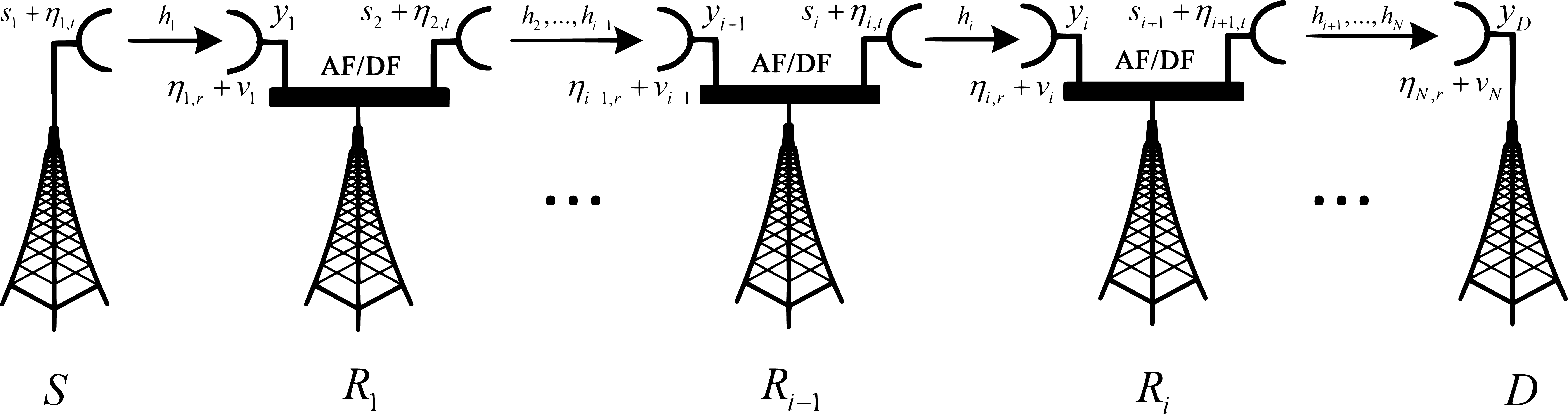}
\caption{{System model of multi-hop relaying with HI}}
\label{fig:Fig. V0}
\end{figure*}
We consider a MH cooperative system over $\bf H$-fading channel using DF and AF protocols. The source node $S$ establishes a link with the destination node $D = R_N$ via multiple relay nodes $R_i $ for $i=1,...,N-1$ (see Fig. 1). {Multiple relays are used to deliver the transmitted signal from source to the final destination.} The received signal at the $i^{\rm th}$ receiver is given by \cite{Bjornson}
\begin{align}
y_i =&\hspace{1mm} \left( h_i \right) ^{\frac{r_i}{2}} \left(s_i+\eta_{i,t} \right)+\eta_{i,r}+v_i,  \quad \forall i=1,...,N,
\label{E1_1}
\end{align}
where $s_1$ and $s_{i+1} \quad \forall i=1,...,N-1$ are respectively the transmitted signals from the source and $i^{\rm th}$ relay, with average signal power $P_i = \mathbb{E}_{s_i}\{|s_i|^2\} \quad \forall i=1,...,N$; $h_i$ is the channel coefficient of the  $i^{\rm th}$ link; $y_N$ is the received signal at the destination; $\eta_{i,t}$ and $\eta_{i,r}$ for $i=1,...,N$ are distortion noises of transmitter and receiver of the $i^{\rm th}$ hop, respectively and are modeled as $\eta_{i,t} \sim \mathcal{CN}(0; \kappa_{i,t}^2 P_i)$, and $\eta_{i,r} \sim \mathcal{CN}(0; \kappa_{i,r}^2 P_i |h_i|^{r_i})$ where $\kappa_{i,t}$, $\kappa_{i,r} \geq 0$ signalize respectively the HI' level in the transmitter and receiver of the $i^{\rm th}$ hop; $v_i \sim \mathcal{CN}(0; \sigma^2_i)$ represents the Gaussian noise for the $i^{\rm th}$ hop's receiver; {$r_i = 1$ and $r_i = 2$ represent the detection mode of HD and DD, respectively}.
Combining the HI at both the transmitter and the receiver and since $\mathbb{E}_{{\eta _{i,t}},{\eta _{i,r}}}\left\{ {{{\left| {\left( h_i \right) ^{\frac{r_i}{2}}{\eta _{i,t}} + {\eta _{i,r}}} \right|}^2}} \right\} = {P_i}{\left| {{h_i}} \right|^{r_i}}\kappa _i^2$ with $\kappa_i^2=\kappa_{i,t}^2+\kappa_{i,r}^2$, we obtain
\begin{align}
y_i =&\hspace{1mm} \left( h_i \right) ^{\frac{r_i}{2}}\left(s_i+\eta_{i} \right)+v_i,  \quad \forall i=1,...,N
\label{E2_1}
\end{align}
where $\eta_i \sim \mathcal{CN}(0; \kappa_i^2 P_i)$ is the aggregated distortion noise for the $i^{\rm th}$ hop. {Note that, when $\kappa_i=\kappa_{i,t}=\kappa_{i,r}=0$, the investigated model corresponds to the ideal transmitter and receiver.} 
 Assuming $\Gamma_i=P_i|h_i|^{r_i}/\sigma^2_i$ for the SNR of $i^{\rm th}$ hop and ${\overline { \Gamma}}_i =P_i\mathbb{E}_{|h_i|^{r_i}}\{|h_i|^{r_i}\}/\sigma^2_i$ for the average SNR of the $i^{\rm th}$ hop, each relay either amplifies or decodes the received signal and forwards it to the next relay for AF and DF relaying, respectively.

The AF relaying protocol essentially amplifies the signal $y_i$ that was received at the relay ($R_i$) to create the transmitted signal $s_{i+1}$ at the relay (i.e., $s_{i+1}=G_i y_i$ for $i=1,...,N-1$), where $G_i$ is the amplification gain of the $i^{\rm th}$ relay.
 For the semi-blind fixed-gain (FG) AF relaying, the input signal at each relay should be normalized, therefore $G_i^2 = \frac{{{P_{i+1}}}}{{{\mathbb{E}_{{s_1},{v_1},{\eta _1},{h_1},...,{v_i},{\eta _i},{h_i}}}\left\{ {y_i^2} \right\}}}$ for $i=1,...,N-1$. For blind FG AF relaying, $G_i$ does not depend on any parameter and is an arbitrary fixed constant. For the channel state information (CSI)-assisted AF relaying, the relay knows the instantaneous information of the fading channel and hence $G_i^2 = \frac{{{P_{i+1}}}}{{{\mathbb{E}_{{s_1},{v_1},{\eta _1},...,{v_i},{\eta _i}}}\left\{ {y_i^2} \right\}}}$ for $i=1,...,N-1$. In addition, the sent signal $s_{i+1}$ at the relay, as specified by the DF relaying protocol, must correspond to the signal's initial value $s_i$.

The source transmits its signal to a relay node. Assuming the FG AF relaying, the $i^{\rm th}$ relay amplifies the signal with a FG $G_i$ and sends it to the next relay. This keeps happening until the data from the source reaches the destination. Therefore, the received signal at the $i^{\rm th}$ receiver $\forall i\in [1,N-1]$ and the destination node ($i=N$) for the AF relaying is expressed as
\begin{align}
{y_i} = \hspace{-1mm}\prod\limits_{j = 1}^i {{G_{j - 1}} h_j  ^{\frac{r_j}{2}}s_1 }+ \hspace{-1mm}\sum\limits_{j = 1}^i \left[{{v_j}}+{{\eta_j} h_j ^{\frac{r_j}{2}}}\right]\hspace{-1mm} \prod\limits_{k = j + 1}^i {{G_{k - 1}} h_k ^{\frac{r_k}{2}}}
\label{E3_1}
\end{align}
for $i=1,...,N$. For the DF relay, the received signal at the $i^{\rm th}$ receiver can be written as
\begin{align}
{y_i} =& \hspace{1mm} \left( h_i \right) ^{\frac{r_i}{2}}\left( s'_i+\eta_i \right)+v_i, \quad \forall i=1,...,N
\label{E4_1}
\end{align}
where $s'_i$ is the retrieved signal at the $i^{\rm th}$ receiver.

The probability density function (PDF) of $i^{\rm th}$ link's instantaneous SNR with H-distribution is defined as  \cite[Eq. (1)]{TVTH}
\begin{align}
{f_{{\Gamma _i}}}\left( \gamma  \right) = \sum_{l_i=1}^{\alpha_i}\frac{{{\rho _{i}}}}{\varrho _i\gamma }{\bf{H}}_{{p_i},{q_i}}^{{m_i},{n_i}}\left[ {\frac{\varrho _i}{\overline \Gamma_i }\gamma \left| {\begin{array}{*{20}{c}}
{\left( {{a_i} ,{A_i}} \right)}\\
{\left( {{b_{i}} ,{B_i}} \right)}
\end{array}} \right.} \right]
\label{pdf_H}
\end{align}
where ${\mathop{\bf H}\nolimits} _{{p_i},{q_i}}^{{m_i},{n_i}}\left[.\right]$ is the $\bf H$-function defined in \cite[Eq. (1.2)]{mathai}, while  $\alpha_i$, $\rho _i$, $\varrho _i$, ${m_i}$, ${n_i}$, ${p_i}$, ${q_i}$, ${\overline \Gamma_i }$, ${a_i}$, ${A_i}$, ${b_i}$, and ${B_i}$ are the parameters of the distribution corresponding to the $i^{\rm th}$ hop.

The cumulative distribution function (CDF) of $i^{\rm th}$ link instantaneous SNR with $\bf H$-distribution by applying \cite[Eqs. (2.53, 2.54)]{mathai} to \eqref{pdf_H} is derived as
\begin{align}
{F_{{\Gamma _i}}}\left( \gamma  \right) = \sum_{l_i=1}^{\alpha_i}\frac{{{\rho _{i}}}}{\varrho _i }{\bf{H}}_{{p_i} + 1,{q_i} + 1}^{{m_i},{n_i} + 1}\left[ { {\frac{\varrho _i}{\overline \Gamma_i }}\gamma \left| {\begin{array}{*{20}{c}}
{\left( {[1,{a_i} ],[1,{A_i}]} \right)}\\
{\left( {[{b_{i}} ,0],[{B_i},1]} \right)}
\end{array}} \right.} \right]
\label{cdf_H}
\end{align}
At high SNRs, when ${\overline \Gamma_i }\gg 1$, the CDF in \eqref{cdf_H} can be approximated by employing \cite[Eqs. (1.8.4, 1.8.5)]{Kilbas} as
\begin{align}
{F_{{\Gamma _i}}}\left( \gamma  \right) \approx \sum_{l_i=1}^{\alpha_i}\sum\limits_{j = 1}^{{m_i}} D_{i,j}{{\left( {\frac{\gamma}{\overline \Gamma_i }} \right)}^{\beta_{i,j}}},\quad \forall i=1,...,N
\label{cdf_H_hsnr}
\end{align}
where ${\beta_{i,j}} \triangleq \frac{b_{i,j}}{B_{i,j}}$ is the $j$-th element of ${\beta_{i}}$, and
\begin{align}
D_{i,j} \triangleq & \hspace{2mm} \frac{{{\rho _i}}}{\varrho _i{{b_{i,j}} }}\frac{{\prod\limits_{k=1,k\ne j}^{{m_i}} {\Gamma \left( {{b_{i,k}} } - \frac{{{{{b_{i,j}} }B_{i,k}}}}{{{B_{i,j}}}} \right) } }}{\prod\limits_{k = {n_i} + 1}^{{p_i}} {\Gamma \left( {{a_{i,k}}- \frac{{{{{b_{i,j}} }A_{i,k}}}}{{{B_{i,j}}}}} \right)}}
\nonumber\\ \times &
\frac{\prod\limits_{k = 1}^{{n_i}} {\Gamma \left( {1 - {a_{i,k}} + \frac{{{{{b_{i,j}} }A_{i,k}}}}{{{B_{i,j}}}}} \right)}}{{\prod\limits_{k = {m_i} + 1}^{{q_i}} {\Gamma \left( {1 - {b_{i,k}}  + \frac{{{{{b_{i,j}} }B_{i,k}}}}{{{B_{i,j}}}}} \right)} }}{{\left( {{\varrho _i} } \right)}^{{\beta_{i,j}} }}
\label{cdf_H_hsnrD}
\end{align}
Some Special cases of $\bf H$-distribution, by dropping the index $i$, are given in Table I, where DGG, EGK, $\boldsymbol\alpha-\boldsymbol\mu$, $\mathcal{F}$, Malaga, mEGG, $N\ast$ Nakagami-$m$  distribution parameters are defined respectively in \cite{TWC2020}, \cite{yilmaz}, \cite{MHTHz}, \cite{MultihopF}, \cite{Malaga}, \cite{EGG}, and \cite{KaragiannidisN}.
\begin{table*}[!ht]
\fontsize{9pt}{9pt}\selectfont
\caption {Some Special cases of $\bf H$-distribution.} 
\centering
\renewcommand{\arraystretch}{2}%
\begin{tabular}{|c||c|c| c| c| c| c|}
\hline
$\bf H$-fading    & Parameters & $\mathbf{\rho}$                   , $\mathbf{\varrho}$             & ${\alpha}$, ${m}$, ${n}$, ${p}$, ${q}$ & ${a}$ ; ${A}$ & ${b}$ ; ${B}$ \\
 \hhline{=======}
DGG+PE         & ${\boldsymbol\alpha_1}, {\boldsymbol\beta _1}, {\Omega _1},r$ & $A_3\varrho$   , ${(\sigma^2_{si}+1)C^{1/v}}{{{}}}$ & 1, $u$, $0$, $r$, $u$ & ${{\boldsymbol{\kappa} _{{3}}}}$ ; ${v^{ - 1}{{\bf{1}}_{{r}}}}$ & ${{\boldsymbol{\kappa} _{{4}}}}$ ; ${v^{ - 1}{{\bf{1}}_{{u}}}}$ \\
\cite{TWC2020}  & ${\boldsymbol\alpha_2}, {\boldsymbol\beta _2}, {\Omega _2}, \xi$ &    &  &  &  \\ \cline{1-7}
EGK            & ${\boldsymbol m}, {\boldsymbol m_s},$         & $\frac{\varrho}{\Gamma (\textit{\textit{\textbf{m}}})\Gamma ({\textit{\textbf{m}}_s})}$ , ${{\textit{\textbf{b}}\textit{\textbf{b}}_s }}{{}}$ & 1, 2, 0, 0, 2 & - ; - & $\left[ {\textit{\textbf{m}},\textit{\textbf{m}}_s} \right]$ ; $\left[ {\frac{2}{\boldsymbol\beta },\frac{2}{\boldsymbol\beta _s}} \right]$ \\
\cite{yilmaz}    & ${\boldsymbol \beta}, {\boldsymbol \beta_s}$ &    &  &  &  \\ \cline{1-7}
$\boldsymbol\alpha$-$\boldsymbol\mu$+PE  & $\boldsymbol\alpha,\boldsymbol\mu , \xi $ &  $\frac{2\varrho}{\boldsymbol\alpha}\frac{A_1}{\left( {{A_2}} \right)^{  \frac{{{\xi ^2}}}{\boldsymbol\alpha}}}$ , ${{{{\left( {{A_2}} \right)}^{\frac{2}{\boldsymbol\alpha}}}}}$ & 1, 2, 0, 0, 2 & $1 + \frac{{{\xi ^2}}}{\boldsymbol\alpha}$ ; $\frac{2}{\boldsymbol\alpha}$ & $[\boldsymbol\mu ,\frac{{{\xi ^2}}}{\boldsymbol\alpha}]$ ; $[\frac{2}{\boldsymbol\alpha},\frac{2}{\boldsymbol\alpha}]$ \\
\cite{MHTHz}  &  &    &  &   &  \\ \cline{1-7}
$\mathcal{F}$  +PE        & $\sigma^2_{\ln S}, \sigma^2_{\ln L},$           &  $\frac{\varrho \xi^2}{ \Gamma(\textit{\textbf{a}}) \Gamma(\textit{\textbf{b}})}$ , $\frac{\textit{\textbf{a}}^r(\textit{\textbf{b}}-1)^{-r}}{ (1+\xi^{-2})^r }$ & 1, 2, 1, 2, 2 & $[1-\textit{\textbf{b}},1+\xi^2]$ ; $[r,r]$ & $[\textit{\textbf{a}},\xi^2]$ ;  $[r,r]$ \\
\cite{MultihopF}     & $r,\xi$  &     &  &  &  \\ \cline{1-7}
Malaga  +PE       & $\boldsymbol\rho, \textit{\textbf{b}}_0, \Omega,\boldsymbol\alpha,\boldsymbol\beta,$            &  $\frac{\varrho r\xi^2\textit{\textbf{A}}\textit{\textbf{b}}_l}{2^r}$ , $(\sigma^2_{si}+1)\textit{\textbf{B}}^r$ & $\boldsymbol\beta$, 3, 0, 1, 3 & $1+\xi^2$ ; $r$ & $[\xi^2,\boldsymbol\alpha,l]$ ; $[r,r,r]$ \\
\cite{Malaga}       & $\phi_A, \phi_B, \xi,r$ &     &  &  &  \\ \cline{1-7}
mEGG   \cite{EGG}    & $\boldsymbol\lambda, \omega, \textit{\textbf{a}}, \textit{\textbf{b}}, \textit{\textbf{c}},r$            &  $\varrho s $ , $\frac{1}{\theta}$ & 2, 1, 0, 0, 1 & - ; - & $\boldsymbol\alpha$ ; $v$ \\ \cline{1-7}
{$N\ast$ Nakagami-$m$}       & {${\textit{\textbf{m}}_i}, \Omega_i, N$}            &  {$\frac{\varrho}{\prod_{i=1}^{N}\Gamma({\textit{\textbf{m}}_i})}  $ , ${\prod_{i=1}^{N}{\textit{\textbf{m}}_i}}$} & {1, $N$, 0, 0, $N$} & {1 ; 1 }&{ ${\textit{\textbf{m}}_1}$,...,${\textit{\textbf{m}}_N}$ ; ${{\bf{1}}_{{N}}}$} \\
{\cite{KaragiannidisN}}  &  &    &  &   &  \\ \cline{1-7}
             \hline
\end{tabular}
 \quad
\label{table1}
\end{table*}
The defined distributions in Table I also contain many distributions. DGG includes K channel, Double Weibull, Gamma-Gamma, and Log-Normal. EGK includes Weibull, Generalized Nakagami-$m$, Generalized Gamma, Generalized-$K$, double Nakagami-$m$. $\boldsymbol\alpha-\boldsymbol\mu$ fading contains Nakagami-$m$, Gamma, Weibull, and Rayleigh. Malaga contains shadowed Rician, Rice-Nakagami, Gamma-Rician. EGG is a special case of mEGG. Finally, double Nakagami-$m$ and triple Nakagami-$m$ are special cases of $N\ast$ Nakagami-$m$.

\section{Performance Analysis of AF Relaying}
{In this section, we derive the E2E instantaneous SNDR statistics of FG, and CSI-assisted $N$-hop AF relaying.
Based on the SNDR expression, we obtain the CDF of dual-hop and MH FG AF relaying with HI in terms of $\bf H$-function. Then, the asymptotic CDF of $N$-hop FG AF relaying with HI, the SNDR ceiling and the diversity order of both blind and semi-blind cases are derived. Additionally, the OP and EC of MH relaying with HI, the capacity ceiling and the BEP of ID are given in closed-form expressions.}
\newline
The SNDR of $N$-hop AF relaying using \eqref{E3_1} can be written as \eqref{SNDR_AFF},
\begin{figure*}
\begin{align}
\Gamma^{X} =\frac{\mathbb{E}_{{{s_1},{v_1},{\eta _1},...,{v_N},{\eta _N}}}\Bigg\{{{{\left| {\prod\limits_{j = 1}^N {{G_{j - 1}}\left( h_j \right) ^{\frac{r_j}{2}}{s_1}} } \right|}^2}}\Bigg\}}{\mathbb{E}_{{{v_1},{\eta _1},...,{v_N},{\eta _N}}}\Bigg\{{{{\left| {\sum\limits_{j = 1}^N {{v_j}} \prod\limits_{k = j + 1}^N {{G_{k - 1}}\left( h_k \right) ^{\frac{r_k}{2}}}  + \sum\limits_{j = 1}^N {{\eta _j}\left( h_j \right) ^{\frac{r_j}{2}}} \prod\limits_{k = j + 1}^N {{G_{k - 1}}} \left( h_k \right) ^{\frac{r_k}{2}}} \right|}^2}}\Bigg\}}
\label{SNDR_AFF}
\end{align}
\hrulefill
\end{figure*}
with $X \in \{F,V\}$ where ``$F$" represents the FG AF relaying and ``$V$" denotes the CSI-assisted or variable gain AF relaying.
\begin{thm}
{The SNDR for $N$-hop FG AF relaying is given by}
\begin{align}
\Gamma^{F} = {\left( {d_1+ \sum\limits_{i = 1}^N {{\lambda _{i + 1}}\prod\limits_{j = 1}^i {\frac{{{C_{{R_{j - 1}}}}}}{{{\Gamma _j}}}} } } \right)^{ - 1}}
\label{SNDR_N}
\end{align}
where $d_i={\lambda _i}-1 \quad \forall i=1,...,N-1$, ${\lambda _i} \triangleq \prod\limits_{j = i}^N {\left( {1 + \kappa _j^2} \right)} $, ${\lambda _{N+1}} =1$, ${C_{R_0}}=1$.
For the FG, ${C_{R_i}} \triangleq \left(1+\kappa_i^2\right)\mathbb{E}\left\{\Gamma_i \right\}+1$ for $i=1,...,N-1$, whereas in the blind case, $C_{R_i}$ is a fixed constant.
\begin{proof}
The expression is derived by putting the expressions for $i^{th}$ relay amplification gain into \eqref{SNDR_AFF}, using the variance of the distortion and receiver noises of all hops, and employing the instantaneous SNR expression for each hop.
See Appendix I for a detailed proof.
\end{proof}
\end{thm}
For the CSI-assisted relaying, ${C_{R_i}} \triangleq \left(1+\kappa_i^2\right)\Gamma_i +1$ $\quad \forall i=1,...,N-1$. Therefore, for the CSI-assisted relaying
\begin{align}
\Gamma^{V}  = \left( \frac{\prod_{i=1}^{N} \left({1+\Gamma'_i}\right)}{\prod_{i=1}^{N} \left({\Gamma_i}\right)}-1 \right)^{-1}
\label{sndr_csi}
\end{align}
where $\Gamma'_i \triangleq \left(1+\kappa_i^2\right)\Gamma_i, \quad \forall i=1,...,N$.\\
Note that, when $\kappa_i=0 \quad \forall i=1,...,N$; \eqref{SNDR_N} and \eqref{sndr_csi} respectively simplify to the SNR of FG and CSI-assisted AF relaying with the ID.

\begin{thm}
\hspace{-1mm}The CDF of dual-hop FG AF relaying with HI over $\bf H$-fading is given by \eqref{dual-hop-HI-exact},
\begin{figure*}
\begin{align}
F_{\Gamma^F}^{{\rm{HI,dual}}}(\gamma ) =&\sum_{l_1=1}^{\alpha_1}\frac{{{\rho _1}}}{{{\varrho _1}}}{\bf{H}}_{{p_1} + 1,{q_1} + 1}^{{m_1},{n_1+ 1}}\left[ {\frac{{{\varrho _1}{\lambda _2} }}{{1 - d_1\gamma }}{\frac{\gamma}{\overline \Gamma_1 }}}{ \left| {\begin{array}{*{20}{c}}
{\left( {[1,{a_1} ],[1,{A_1}]} \right)}\\
{\left( {[{b_1} ,0],[{B_1},1]} \right)}
\end{array}} \right.} \right]+\sum_{l_1=1}^{\alpha_1}\sum_{l_2=1}^{\alpha_2}\frac{{{\rho _1}{\rho _{2}}}}{{{\varrho _1}{\varrho _2}}}
\nonumber\\
 \times&
{\bf{H}}_{1,0:{q_1},{p_1}+1 : {{p_2}}  + 1, {{q_2}}  + 2}^{0,1:{n_1},{m_1}: {{m_2}}  + 1, {{n_2}}  + 1}\hspace{-2mm}\left[ \hspace{-2mm}{\begin{array}{*{20}{c}}
{{{ \frac{{1 - d_1\gamma }}{{\varrho _1}{\lambda _2} } }{\frac{\overline \Gamma_1 }{\gamma}}}}\\
{\frac{{{C_{{R_1}}} }}{\lambda_2}}{\frac{\varrho _2}{\overline \Gamma_2 }}
\end{array}\hspace{-3mm}\left|\hspace{-2mm} {\begin{array}{*{20}{c}}
{\left( {1;1,1} \right):\left( {\left[ {{1-b_1} } \right],\left[ {{B_1}} \right]} \right);\left( {[1,{a_2} ],[1,{A_2}]} \right)}\\
{ - :\left( {\left[ {{1-a_1},0 } \right],\left[ {{A_1}},1 \right]} \right);\left( {[1,{b_2},0],[1,{B_2},1]} \right)}
\end{array}} \right.}\hspace{-3mm} \right]
\label{dual-hop-HI-exact}
\end{align}
\hrulefill
\end{figure*}
where ${\mathbf{H}} _{p_1,q_1:p_2,q_2:p_3,q_3}^{0,n_1:m_2,n_2,m_3,n_3} \left[.\right]$ is bivariate $\bf H$-function defined in \cite[Eq. (2.57)]{mathai}.
The CDF in \eqref{dual-hop-HI-exact} can be approximated as in \eqref{2-hop-AF-HI}.
\begin{figure*}
\begin{align}
F_{{\Gamma ^F}}^{{\rm{HI,dual}}}(\gamma ) \approx&\sum_{l_1=1}^{\alpha_1}
\frac{{{\rho _1}}}{{{\varrho _1}}}{\bf{H}}_{{p_1} + 1,{q_1} + 1}^{{m_1},{n_1} + 1}\left[ {{\frac{{{\varrho _1}{\lambda _2} }}{{1 - d_1\gamma }}{\frac{\gamma}{\overline \Gamma_1 }}}\left| {\begin{array}{*{20}{c}}
{\left( {[1,{a_1}],[1,{A_1}]} \right)}\\
{\left( {[{b_1},0],[{B_1},1]} \right)}
\end{array}} \right.} \right]
\nonumber\\
+ &\sum_{l_1=1}^{\alpha_1}\sum_{l_2=1}^{\alpha_2}\frac{{{\rho _1}{\rho _{2}}}}{{{\varrho _1}{\varrho _2}}}{\bf{H}}_{{p_1} + {p_2} + 1,{q_1} + {q_2} + 1}^{{m_1} + {m_2},{n_1} + {n_2} + 1}\left[ {{\frac{{{\varrho _1}{\varrho _2}{C_{{R_1}}} }}{{1 - d_1\gamma }}}{\frac{\gamma}{\overline \Gamma_1 \overline \Gamma_2 }}\left| {\begin{array}{*{20}{c}}
{\left[ {1,a_1,a_2} \right];\left[ {1,{A_1},{A_2}} \right]}\\
{\left[ {b_1,b_2,0} \right];\left[ {{B_1},{B_2},1} \right]}
\end{array}} \right.} \right]
\label{2-hop-AF-HI}
\end{align}
\hrulefill
\end{figure*}
\begin{proof}
See Appendix II.
\end{proof}
\end{thm}

 { {\bf Special cases:}} It can be shown that when we set $\textit{\textbf{m}}_s\rightarrow \infty, \boldsymbol\beta=\boldsymbol\beta_s=2$ in EGK fading of Table I, the CDF in \eqref{2-hop-AF-HI} simplifies to \cite[Eq. (27)]{Bjornson} earlier obtained for dual-hop AF relaying HI systems over Nakagami-$m$ channels.
 Moreover, it can be shown that if we set $r_1=r_2=N=2$ and consider $\boldsymbol\alpha$-$\boldsymbol\mu$+PE fading for the THz link and DGG with PE for the FSO link, the CDF in \eqref{dual-hop-HI-exact} reduces to \cite[Eq. (34)]{Jiayi} earlier obtained for dual-hop THz-FSO relaying systems with HI.
 It can be shown that if we consider Malaga+PE fading for the FSO link, the CDF in \eqref{2-hop-AF-HI} simplifies to \cite[Eq. (27)]{Balti} earlier obtained for dual-hop RF-FSO relaying systems with HI over Rayleigh-Malaga channels with partial relay selection and outdated CSI.
 In another special case that considers the ID in all nodes, i.e., $\kappa_1=\kappa_2=0$, over Nakagami-$m$, i.e., $\textit{\textbf{m}}_s\rightarrow \infty, \boldsymbol\beta=\boldsymbol\beta_s=2$, and Gamma-Gamma, i.e., ${\boldsymbol\alpha_1}_i={\Omega_1}_i={\boldsymbol\alpha_2}_i={\Omega_2}_i=1$, the CDF expression in \eqref{2-hop-AF-HI} is reduced to \cite[Eq. (8)]{Zedini-FSO}.
For ID, the CDF is derived by setting $\lambda_1=\lambda_2=1$ in \eqref{dual-hop-HI-exact} and \eqref{2-hop-AF-HI}.

\begin{thm}
The CDF of $N$-hop FG AF relaying with HI over $\bf H$-fading is given by \eqref{N-hop-AF-exact},
\begin{figure*}
\begin{align}
F_{\Gamma^F}^{\rm HI}(\gamma ) \approx &\sum_{l_1=1}^{\alpha_1}\frac{\rho _1}{\varrho _1}{\bf{H}}_{{p_1} + 1,{q_1} + 1}^{{m_1},{n_1+ 1}}\left[ {{\frac{{{\varrho _1}{\Lambda _1} }}{{1 - d_1\gamma }}{\frac{\gamma}{\overline \Gamma_1 }}} \left| {\begin{array}{*{20}{c}}
{\left( {[1,{a_1} ],[1,{A_1}]} \right)}\\
{\left( {[{b_1} ,0],[{B_1},1]} \right)}
\end{array}} \right.} \right]
+ \sum\limits_{i = 2}^N \sum_{l_2=1}^{\alpha_2}...\sum_{l_i=1}^{\alpha_i}{\prod\limits_{j = 1}^i {\frac{\rho _j}{\varrho _j}} }
\nonumber\\
\times &{\bf{H}}_{1,0:{q_1},{p_1}+1 :\sum\limits_{j = 2}^i {{p_j}}  + 1,\sum\limits_{j = 2}^i {{q_j}}  + 2}^{0,1:{n_1},{m_1}:\sum\limits_{j = 2}^i {{m_j}}  + 1,\sum\limits_{j = 2}^i {{n_j}}  + 1}\left[ \hspace{-2mm}{\begin{array}{*{20}{c}}
{{{\frac {\left({{1 - d_1\gamma }}\right)}{{{\lambda_{1,2}}\varrho _1} } }{\frac{\overline \Gamma_1 }{\gamma}}}}\\
\frac{{\Lambda _i}}{\lambda_{1,2}}{\prod\limits_{j = 2}^{i} \frac {{C_{{R_{j-1}}}}\varrho _j}{{\overline \Gamma_j}} }
\end{array}\hspace{-3mm}\left|\hspace{-2mm} {\begin{array}{*{20}{c}}
{\left( {1;1,1} \right):\left( {\left[ {{1-b_1} } \right],\left[ {{B_1}} \right]} \right);\left( {[1,a_{2,...,i} ],[1,A_{2,...,i}]} \right)}\\
{ - :\left( {\left[ {{1-a_1},0 } \right],\left[ {{A_1}},1 \right]} \right);\left( {[1,b_{2,...,i} ,0],[1,B_{2,...,i},1]} \right)}
\end{array}} \right.}\hspace{-3mm} \right]
\label{N-hop-AF-exact}
\end{align}
\hrulefill
\end{figure*}
and approximated as \eqref{N-hop-AF-HI},
\begin{figure*}
\begin{align}
F_{{\Gamma ^F}}^{{\rm HI}}(\gamma ) \approx \sum\limits_{i = 1}^N \sum_{l_1=1}^{\alpha_1}...\sum_{l_i=1}^{\alpha_i}{\prod\limits_{j = 1}^i {\frac{\rho _j}{\varrho _j}} } {\bf{H}}_{\sum\limits_{j = 1}^i {{p_j}}  + 1,\sum\limits_{j = 1}^i {{q_j}}  + 1}^{\sum\limits_{j = 1}^i {{m_j}} ,\sum\limits_{j = 1}^i {{n_j}}  + 1}\left[ {\frac{{\prod\limits_{j = 1}^i \left( {{C_{{R_{j - 1}}}}}\varrho _j\right) {\Lambda _i}\gamma }}{\left({{1 - d_1\gamma }}\right)\prod\limits_{j = 1}^i {{\overline \Gamma_j }}}\left| {\begin{array}{*{20}{c}}
{\left( {[1,a_{1,...,i}],[1,A_{1,...,i}]} \right)}\\
{\left( {[b_{1,...,i} ,0],[B_{1,...,i},1]} \right)}
\end{array}} \right.} \right]
\label{N-hop-AF-HI}
\end{align}
\hrulefill
\end{figure*}
where
${\Lambda _i} \triangleq \left\{ {\begin{array}{*{20}{c}}
{{\lambda _{i,i + 1}}\hspace{3mm} {\rm if} \hspace{2mm}  i = 1,...,N - 2}\\
\hspace{-2mm}{{\lambda _{i + 1}}\hspace{5mm} {\rm if} \hspace{3mm}  i = N - 1,N{\rm{    }}}
\end{array}} \right.$,
and ${\lambda _{i,i + 1}} \triangleq {\lambda _{i + 1}} + {C_{{R_i}}}\left( {1 - {\lambda _{i + 1}}} \right), \forall i=1,...,N-1$,
$a_{i,...,j} \triangleq {a_i} ,...,{a_j}$, $A_{i,...,j} \triangleq {A_i} ,...,{A_j}$, $b_{i,...,j} \triangleq {b_i} ,...,{b_j}$, $B_{i,...,j} \triangleq {B_i} ,...,{B_j}$.

\begin{proof}
See Appendix III.
\end{proof}
\end{thm}
Note that, the CDF expressions of the ID cases can be obtained by setting $\kappa_i=0, \quad \forall i=1,...,N$ in Theorem 2.
Assuming $\kappa_i=0, \quad \forall i=1,...,N$ and DGG plus PE and EGK fading respectively for the FSO and RF channels, the CDF in \eqref{N-hop-AF-HI} respectively simplifies to \cite[Eq. (24)]{TWC2020} and \cite[Eqs. (25, 28, 32)]{jocn2022} earlier obtained for MH FSO, MH RF, MH mixed FSO-RF, and MH mixed RF-FSO relaying systems.

\subsection{Asymptotic Analysis of AF Relaying}
In this section, we obtain the E2E instantaneous SNDR statistics of FG AF relaying. In contrast with the expression in \eqref{N-hop-AF-HI} which depends on a complicated $\bf H$-function, we derive the closed-form expression of CDF as it includes finite summation of some simple elementary functions.
To deeply explore on the influence of system parameters, such as number of hops and fading parameters, on the system performance.
More importantly, we estimate the diversity order by using the slope of CDF against SNR curve in a logarithmic scale.

\begin{thm}
At high SNRs when ${{{\overline \Gamma }_i}}\gg 1$ $\forall i=1,...,N$, the asymptotic CDF of MH AF relaying with HI is given by
\begin{align}
{F_{{\Gamma ^F}}^{\rm HI,\infty}}(\gamma) \approx  \sum_{l_1=1}^{\alpha_1}\sum\limits_{{i_1} = 1}^{{m_1}} {\frac{{{{D_{1,{i_1}}}}{A_N}}}{{{{\left( {1 - {d_1}\gamma } \right)}^{{\beta _{1,{i_1}}}}}}}} {\left( {\frac{\gamma }{{{{\overline \Gamma }_1}}}} \right)^{^{{\beta _{1,i_1}}}}}
\label{CDF_HI_N_hsnr}
\end{align}
where ${A_N}$ is defined in \eqref{cdf_H_hsnr0}, while $_p{F_q}(·)$ is the Hypergeometric function defined in \cite[Eq. (9.111)]{Gradshteyn:2007}.
\begin{figure*}
\begin{align}
{A_N} \triangleq & \hspace{2mm}
  \lambda _{2}^{{\beta _{1,{i_1}}}}+
\sum\limits_{n = 2}^{N - 1}\sum_{l_2=1}^{\alpha_2}...\sum_{l_n=1}^{\alpha_n} {\sum\limits_{{i_2} = 1}^{{m_2}} {...\sum\limits_{{i_n} = 1}^{{m_n}} {\prod\limits_{j = 2}^{n - 1} {{\beta _{j,{i_j}}}{D _{j,{i_j}}}} } D_{n,{i_n}}} } \lambda _{n + 1}^{{\beta _{n,{i_n}}}}\prod\limits_{j = 2}^n {\frac{{{{\left( {{\lambda _{j - 1,j}}} \right)}^{{\beta _{j - 1,{i_{j - 1}}}}}}}}{{d_{j}^{{\beta _{j,{i_j}}}}}\left({\overline \Gamma_j }\right)^{\beta _{j,i_j}}}}
 \nonumber\\
\times &{\left( { - 1} \right)^{ { - \sum\limits_{j = 2}^n {{\beta _{j,{i_j}}}} }}}\prod\limits_{j = 2}^n {\frac{1}{{{\beta _{j,{i_j}}} - {\beta _{j - 1,{i_{j - 1}}}}}}}\prod\limits_{j = 2}^n {{}_2{F_1}\left( {{\beta _{j,{i_j}}},1;{\beta _{j,{i_j}}} - {\beta _{j - 1,{i_{j - 1}}}} + 1,1 + \frac{{{\lambda _{j - 1,j}}}}{{{d_{j}}{C_{{R_{j - 1}}}}}}} \right)}
 \nonumber\\
 + &\sum_{l_2=1}^{\alpha_2}...\sum_{l_N=1}^{\alpha_N}\sum\limits_{{i_2} = 1}^{{m_2}} {...\sum\limits_{{i_N} = 1}^{{m_N}}}{}\prod\limits_{j = 2}^{N - 1}\frac{{{\beta _{j,{i_j}}}}{{D _{j,{i_j}}}}D _{N,{i_N}}C_{{R_{N - 1}}}^{{\beta _{N,{i_N}}}}}{\lambda _N^{{\beta _{N,{i_N}}}-{\beta _{N - 1,{i_{N - 1}}}} }}
\frac{{\Gamma \left( {1 - {\beta _{N,{i_N}}}} \right)\Gamma \left( {{\beta _{N,{i_N}}} - {\beta _{N - 1,{i_{N - 1}}}}} \right)}}{{\Gamma \left( {1 - {\beta _{N - 1,{i_{N - 1}}}}} \right)}\prod\limits_{j = 2}^{N - 1} {\frac{1}{{{\beta _{j,{i_j}}} - {\beta _{j - 1,{i_{j - 1}}}}}}}}
 \nonumber\\
\times &
\frac{{\left( { - 1} \right)^{ { - \sum\limits_{j = 2}^{N - 1} {{\beta _{j,{i_j}}}} }}}}{\prod\limits_{j = 2}^{N } \left({\overline \Gamma_j }\right)^{\beta _{j,i_j}}}\prod\limits_{j = 2}^{N - 1} {\frac{{{{\left( {{\lambda _{j - 1,j}}} \right)}^{{\beta _{j - 1,{i_{j - 1}}}}}}}}{{d_{j}^{{\beta _{j,{i_j}}}}}}}  \prod\limits_{j = 2}^{N - 1} {{}_2{F_1}\left( {{\beta _{j,{i_j}}},1;{\beta _{j,{i_j}}} - {\beta _{j - 1,{i_{j - 1}}}} + 1,1 + \frac{{{\lambda _{j - 1,j}}}}{{{d_{j}}{C_{{R_{j - 1}}}}}}} \right)}
\label{cdf_H_hsnr0}
\end{align}
\hrulefill
\end{figure*}
\begin{proof}
See Appendix IV.
\end{proof}
\end{thm}
 The diversity order of AF relaying is derived by using the asymptotic CDF expressions in \eqref{CDF_HI_N_hsnr} in the high SNR regime.
 Mathematically speaking, if we increase all links' SNRs with no bound in \eqref{SNDR_N} for AF relaying when $ \gamma \leq \frac{1}{\lambda_1-1}$, we can write
\begin{align}
{{\Gamma ^{F,\infty }}} \triangleq &\mathop {\lim }\limits_{{{\overline \Gamma  }_1},...,{{\overline \Gamma  }_N} \to \infty } \Gamma^{F} = \left({{\prod\limits_{i = 1}^N {\left( {1 + \kappa _i^2} \right) - 1} }}\right)^{-1}
\nonumber\\
=& \frac{1}{{\sum\limits_{i = 1}^N {\frac{1}{{i!}}\underbrace { \sum\limits_{{j_1} = 1}^N {...\sum\limits_{{j_i} = 1}^N {} } }_{{j_1} \ne {j_2} \ne ... \ne {j_i}}\prod\limits_{t = 1}^i {\kappa _{{j_t}}^2} } }}
\label{ceiling_AF}
\end{align}
An SNDR ceiling appears at low outage regime. This can significantly limit the performance of AF relaying systems. Therefore, a CDF floor occurs. In fact, when all links' SNRs grow with bound (i.e., by taking the limit ${{\overline \Gamma  }_1}\rightarrow \infty$ with ${{\overline \Gamma  }_j}=\mu_j {{\overline \Gamma  }_1} \quad \forall j =2,...,N$), the E2E SNDR in \eqref{SNDR_N} converges to $\frac{1}{\lambda_1-1}$. This is different from the ID, where $\mathop{\lim }\limits_{{{\overline \Gamma  }_i} \to \infty } \Gamma^{F}\rightarrow \infty$, since in the ID $d_1$ in \eqref{SNDR_N} is equal to zero.
Thus, an SNDR ceiling effect is observed in the low outage regime, which imposes a limit on the performance of AF relaying system, where for the threshold SNR lower than
the ceiling, the CDF becomes zero with increasing SNR. Note that the SNDR ceiling in \eqref{ceiling_AF} does not depend on the fading distributions of the hops and is inversely proportional to the HI level of all hops' transceivers. When we assume an equal HI level for all hops (i.e., $\kappa_1=...=\kappa_N=\kappa$), using \eqref{ceiling_AF} the necessary and not sufficient condition on the HI level for the AF is given by
\begin{align}
\kappa \le \sqrt{\sqrt[\mathlarger{\mathlarger{N}}]{{1 + \frac{1}{\gamma}}} - 1}
\end{align}
On this basis, the diversity order of system with HI for $ \gamma < \frac{1}{\lambda_1-1}$ is estimated.
Assuming $\beta'_{i} \triangleq \left[\beta_{i,1},...,\beta_{i,m_i}\right]$, utilizing \cite[Eq. (1)]{Wang} and using \eqref{CDF_HI_N_hsnr} when ${{\overline \Gamma  }_i}\gg1 \quad \forall i=1,...,N$, the diversity orders for AF relays with blind (B) relaying  and semi-blind (SB) relaying are respectively given as
\begin{align}
&{G_d^{ F, \rm B}} = \min \Big\{{{{\beta'_{1} }}},2{{{\beta'_{2} }}},...,N{{{\beta'_{N} }}}\Big\}
\label{Gd_AF}
\\
&{G_d^{ F, \rm SB}} = \min \Big\{{{{\beta'_{1} }}},{{{\beta'_{2} }}},...,{{{\beta'_{N} }}}\Big\}
 \label{Gd_AF2}
\end{align}
The diversity order is a function of the number of links and the parameters of the $\bf H$-distribution of $i^{\rm th}$ link.
As anticipated, for the $i^{\rm th}$ link in \eqref{Gd_AF}, the index $i$ $\forall i=1,...,N$ appears as a weighting coefficient on $\beta_{i,j}$ for $j=1,...,m_i$, while there is no index $i$ in \eqref{Gd_AF2}.
\vspace{-2mm}
 \subsection{Outage Probability of AF Relaying}
 The OP  is the probability that the SNR of the E2E system falls below a predetermined threshold SNDR ${{\gamma _{th}}}$, as given by
 \begin{align}
 {P^{\rm HI, F}_{out}}\left( {{\gamma _{th}}} \right) =Pr\left({\Gamma^F }< {{\gamma _{th}}} \right)= {F_{{\Gamma^F}}^{\rm HI}}\left( {{\gamma _{th}}} \right)
\end{align}
 \subsection{Average Bit Error Probability of AF Relaying}
The average BEP of a variety of modulations can be written in terms of the CDF of ${\Gamma }$ as given by \cite{TCOM2019}
 \begin{align}
{\overline P_e} = \frac{{\boldsymbol\delta}}{{2\Gamma \left( p \right)}}\sum\limits_{k = 1}^\mathbf{n} q_k^p{\int\limits_0^\infty {{\gamma ^{p - 1}}} } {\rm{exp}}\left( { - {q_k}\gamma } \right){F^X_\Gamma }(\gamma )d\gamma,
\label{Pe}
\end{align}
where $\boldsymbol\delta $, $p$, ${q_k}$ and $\mathbf{n}$ indicate various modulation techniques as given in \cite{TCOM2019}.
{By substituting \eqref{N-hop-AF-HI} or \eqref{CDF_HI_N_hsnr} into \eqref{Pe}, the obtained integral on $\gamma$ cannot be solved. Therefore, the BEP for the HI cannot be obtained, analytically.} For the ID, we can use the CDF expression in \eqref{N-hop-AF-HI}. By substituting \eqref{N-hop-AF-HI}, when $\kappa_i=0$ for $i=1,...,N$, into \eqref{Pe} and using \cite[Eq. (3.382.4)]{Gradshteyn:2007}, we can obtain the BEP for FG AF with ID as \eqref{Pe_HVLC}.
\begin{figure*}
\begin{align}
&{\overline P_e^{F, {\rm ID}}} = \frac{\boldsymbol\delta}{2\Gamma \left( p \right)} \sum\limits_{k = 1}^{\mathbf{n}}
\sum\limits_{i = 1}^N \sum_{l_1=1}^{\alpha_1}...\sum_{l_i=1}^{\alpha_i}{\prod\limits_{j = 1}^i {\frac{\rho _j}{\varrho _j} }}
 {\bf{H}}_{\sum\limits_{j = 1}^i {{p_j}}  + 2,\sum\limits_{j = 1}^i {{q_j}} + 1}^{\sum\limits_{j = 1}^i {{m_j}} ,\sum\limits_{j = 1}^i {{n_j}}  + 2}\hspace{-1mm}\left[\hspace{-1mm} { {\frac{\prod\limits_{j = 1}^i{C_{{R_{j - 1}}}\varrho _j}}{q_k \prod\limits_{j = 1}^i{\overline \Gamma_j }}}  \hspace{-1mm}\left|\hspace{-1mm} {\begin{array}{*{20}{c}}
{\left( {[1-p,1,a_{1,...,i}],[1,1,A_{1,...,i}]} \right)}\\
{\left( {[b_{1,...,i} ,0],[B_{1,...,i},1]} \right)}
\end{array}} \right.}\hspace{-3mm} \right]
\label{Pe_HVLC}
\end{align}
\hrulefill
\end{figure*}

\subsection{Ergodic Capacity of AF Relaying}
The EC for a MH relaying system is defined as \cite{TCOM2019}
 \begin{align}
\overline C^X =&\hspace{1mm}\frac{1}{N} \mathbb{E}_{\Gamma^X}\left[ {\log_2\left( {1 + c\Gamma^X } \right)} \right]\nonumber\\
 =&\hspace{1mm}\frac{c}{{N\ln\left( 2 \right)}}\int\limits_0^\infty {\frac{{1-F_{\Gamma^X} \left( \gamma \right)}}{{1 + c\gamma }}} d\gamma
\label{cap}
\end{align}
where $c = e/2\pi, 1 $ respectively for the DD and HD methods \cite[Eq. (7.43)]{Advanced}. Note that, the factor $\frac{1}{N}$ accounts for the reason that the whole
communication needs $N$ time slots in a $N$-hop relaying.
Due to mathematical intractability, we cannot apply \eqref{N-hop-AF-HI} directly to \eqref{cap}. Therefore, we use an approximation which is based on the Jensen inequality as in \cite[Eq. (35)]{Bjornson}.
Thus, by substituting \eqref{SNDR_N} in \eqref{cap} and using the Jensen inequality, we obtain
{
\begin{align}
\overline C^{F, {\rm{HI}}} \leq &
 \frac{c}{N}{\log _2}\left[ {\prod\limits_{i = 1}^N {{{\mathbb{E}}}\{ {\Gamma _i}\} } } \right] - \frac{c}{N}{\log _2}\Bigg[ {\prod\limits_{i = 1}^N {\left( {1 + \kappa _i^2} \right)} {{\mathbb{E}}}\{ {\Gamma _i}\} }
 \nonumber\\
 - &\prod\limits_{i = 1}^N {{{\mathbb{E}}}\{ {\Gamma _i}\} }  + \sum\limits_{i = 1}^N \prod\limits_{j = 1}^{i - 1} {{C_{{R_j}}}} \prod\limits_{j = i + 1}^N {\left( {1 + \kappa _j^2} \right)} {{\mathbb{E}}}\{ {\Gamma _j} \}  \Bigg]
\label{EC_F_HI}
\end{align}}
where ${\mathbb{E}\{\Gamma _i\}} \quad \forall i=1,...,N$ is obtained by applying \cite[Eq. (2.8)]{mathai} on \eqref{pdf_H} as
\begin{align}
{\mathbb{E}\{\Gamma _i\}}=\sum_{l_i=1}^{\alpha_i}{\frac{\rho _i}{\varrho _i^2}}
\frac{{\prod\limits_{j = 1}^{{m_i}} {\Gamma \left( {{b_j} + {B_j}} \right)} \prod\limits_{j = 1}^{{n_i}} {\Gamma \left( {1 - {a_j} - {A_j}} \right)} }}{{\prod\limits_{j = {m_i} + 1}^{{q_i}} {\Gamma \left( {1 - {b_j} - {B_j}} \right)} \prod\limits_{j = {n_i} + 1}^{{p_i}} {\Gamma \left( {{a_j} + {A_j}} \right)} }}
\end{align}
By applying {\eqref{ceiling_AF}} to {\eqref{cap}}, we obtain SNDR capacity ceiling of AF relaying as a function of the level of HI as
\begin{align}
{\overline C ^{F,\infty }} = \frac{1}{N}{\log _2}\left( {1 + c{\Gamma ^{F,\infty }}} \right)
\end{align}
For the ID, we can apply \eqref{N-hop-AF-exact} directly to \eqref{cap}. Therefore, by putting \eqref{N-hop-AF-exact}, when $\kappa_i=0$ for $i=1,...,N$, in \eqref{cap} and using \cite[Eq. (1.43)]{mathai}, \cite[Eqs. (2.141.1, 3.194.5)]{Gradshteyn:2007}, we obtain the EC of FG AF as \eqref{EC_F_ID}.
\begin{figure*}
  \begin{align}
\overline C^{F, {\rm{ID}}} \approx & \sum_{l_1=1}^{\alpha_1}\frac{{{\rho _1}}}{{N\ln \left( 2 \right)\varrho _1}}{\rm{ }}{\bf{H}}_{{p_1} + 2,{q_1} + 2}^{{m_1} + 2,{n_1} + 1}\left[ {\frac{{{\varrho _1}}}{c}\left| {\begin{array}{*{20}{c}}
{\left( {[0,{a_1},1 ],[1,{A_1},1]} \right)}\\
{\left( {[0,0,{b_1} ],[1,1,{B_1}]} \right)}
\end{array}} \right.} \right]+\frac{1}{{N\ln \left( 2 \right)}}\sum\limits_{i = 2}^N \sum_{l_2=1}^{\alpha_2}...\sum_{l_i=1}^{\alpha_i} {\prod\limits_{j = 1}^i {\frac{\rho _j}{\varrho _j}} }
\nonumber\\
\times &  {\bf{H}}_{1,0:{q_1}+1,{p_1}+2  :\sum\limits_{j = 2}^i {{p_j}}  + 1,\sum\limits_{j = 2}^i {{q_j}}  + 2}^{0,1:{n_1}+1,{m_1}+1:\sum\limits_{j = 2}^i {{m_j}}  + 1,\sum\limits_{j = 2}^i {{n_j}}  + 1}
 \hspace{-2mm} \left[\hspace{-2mm} {\begin{array}{*{20}{c}}
{{\frac{c}{{{\varrho _1} }{\overline \Gamma_1 }}}}\\
{\prod\limits_{j = 2}^{i }\frac {{C_{{R_{j-1}}}}\varrho _j}{{\overline \Gamma_j }} }
\end{array}\hspace{-3mm} \left|\hspace{-2mm}  {\begin{array}{*{20}{c}}
{\left( {1;1,1} \right):\left( {\left[ 1,{{1-b_1} } \right],\left[ 1,{{B_1}} \right]} \right);\left( {[1,a_{2,...,i} ],[1,A_{2,...,i}]} \right)}\\
{ - :\left( {\left[1, {{1-a_1} ,0} \right],\left[ 1,{{A_1},1} \right]} \right);\left( {[1,b_{2,...,i} ,0],[1,B_{2,...,i},1]} \right)}
\end{array}} \right.}\hspace{-2mm}  \right]
\label{EC_F_ID}
\end{align}
\hrulefill
\end{figure*}

\section{Performance Analysis of DF Relaying}
{In this section, we obtain the E2E instantaneous SNDR statistics of $N$-hop DF relaying.
Based on the SNDR expression, we obtain the CDF of MH DF relaying with HI in terms of $\bf H$-function. Then, the asymptotic CDF of $N$-hop DF relaying with HI, the SNDR ceiling and the diversity order are derived. Additionally, the OP and EC of MH relaying with HI, the capacity ceiling and the BEP of ID are given in closed-form expressions.}

If the $i^{\rm th}$ relay can decode the signal, the effective SNDR is the minimum of the SNDRs between $S-R_1$, $R_1-R_2$,...,$R_{N-1}-D$.
Under the assumption of DF relaying where the destination knows the statistics of the fading and distortion noises of all hops, using \eqref{E4_1}, the E2E SNDR can be expressed as
\begin{align}
{\Gamma }^{D}= \underset{i\in \left\{ 1,...,N \right\}}{\mathop{\min }}\,\left(\frac{\Gamma _{i}}{{\kappa^2 _{i}}{\Gamma _{i}}+1}\right)
\label{sndr_df}
\end{align}
where $D$ represents the DF relaying.
Since the fading and distortion noises of all links are independent, the CDF of $N$-hop DF relaying with HI over $\bf H$-function is given by
\begin{align}
F_{{\Gamma ^D}}^{{\rm{HI}}}(\gamma) =
 {1 - \prod\limits_{i = 1}^N {\left[ {1 - {F_{{\Gamma _i}}}\left( {\frac{\gamma}{{1 - \kappa _i^2\gamma}}} \right)} \right] ,  \hspace{1cm}   \gamma \le \frac{1}{\delta^2 }{\rm{ }}} },
 \label{N-hop-DF-HI}
\end{align}
and $F_{{\Gamma ^D}}^{{\rm{HI}}}(\gamma)=1$, when ${\gamma > \frac{1}{\delta^2 }}$, where $\delta  = \max \left( {\kappa _1,...,\kappa _N} \right)$ and the CDF of $i^{\rm th}$ link is defined earlier in \eqref{cdf_H}.
When  $\gamma \le \frac{1}{\delta^2 }$, the CDF in \eqref{N-hop-DF-HI}  can be approximated as
\begin{align}
&F_{{\Gamma ^D}}^{{\rm{HI}}}(\gamma) \approx\sum\limits_{i = 1}^N  { {F_{{\Gamma _i}}}\left( {\frac{\gamma}{{1 - \kappa _i^2\gamma}}} \right)}
\nonumber\\
&= \sum\limits_{i = 1}^N \sum_{l_i=1}^{\alpha_i} \frac{\rho _i}{\varrho _i}{\bf{H}}_{{p_i} + 1,{q_i} + 1}^{{m_i},{n_i} + 1}\left[ { {\frac{{\varrho _i}}{{1 - \kappa _i^2\gamma}}}\frac{\gamma}{{{{\overline \Gamma  }_i}}}  \left| {\begin{array}{*{20}{c}}
{\left( {[1,{a_i} ],[1,{A_i}]} \right)}\\
{\left( {[{b_i} ,0],[{B_i},1]} \right)}
\end{array}} \right.} \right]
\label{N-hop-DF-HI-approx}
\end{align}
by considering the dominant CDF terms of {\eqref{N-hop-DF-HI}}, where this tight approximation becomes exact at high SNRs. This expression is a tight approximation for the CDF of MH hardware impaired CSI-assisted AF relaying. Note that, when $\kappa_i = 0$ for $i = 1,...,N$, \eqref{N-hop-DF-HI-approx} simplifies to the CDF of MH DF relaying with ID.

\subsection{Asymptotic Analysis of DF Relaying}
For the DF relaying, by applying \eqref{cdf_H_hsnr} to \eqref{N-hop-DF-HI}, the E2E CDF at high SNRs is given as
\begin{align}
F_{{\Gamma ^D}}^{{\rm{HI}}}(\gamma) \approx  &
\sum\limits_{i = 1}^N \sum_{l_i=1}^{\alpha_i} \sum\limits_{j = 1}^{m_i}\frac{{{D_{i,j}}}}{{{{\left( {1 - \kappa _i^2\gamma} \right)}^{{\beta _{i,j}}}}}}{\left( {\frac{\gamma}{{{{\overline \Gamma  }_i}}}} \right)^{{\beta _{i,j}}}}
\label{N-hop-DF-HI2}
\end{align}

If we increase all links' SNRs with no bound in \eqref{sndr_df} for DF relaying, an SNDR ceiling appears in the low outage regime. This can significantly limit the performance of DF relaying systems. Therefore, a CDF floor occurs.
In fact, when the SNRs of all links grow with no bound (i.e., by taking the limit ${{\overline \Gamma  }_1}\rightarrow \infty$ with ${{\overline \Gamma  }_j}=\mu_j {{\overline \Gamma  }_1} \quad \forall j =2,...,N$), the E2E SNDR in \eqref{sndr_df} converges to $\frac{1}{\delta ^2}$ as
\begin{align}
&{\Gamma ^{D,\infty }} \triangleq \mathop {\lim }\limits_{{{\overline \Gamma  }_1},...,{{\overline \Gamma  }_N} \to \infty } \Gamma ^{D} = \frac{1}{\max \left( {\kappa^2_1,...,\kappa^2_N} \right)}
\label{ceiling_DF}
\end{align}
When the HI level of all hops is equal (i.e., $\kappa_1=...=\kappa_N=\kappa$), using \eqref{ceiling_DF}, the necessary and not sufficient condition on the HI for the DF relaying is given by
\begin{align}
\kappa \le \sqrt{\frac{1}{\gamma}} \quad \forall i=1,...,N
\end{align}
Therefore, assuming $\kappa_1=...=\kappa_N=\kappa$, the SNDR ceiling of the DF relaying in {\eqref{ceiling_DF}} is almost $N$ times the SNDR ceiling of the AF relaying given in {\eqref{ceiling_AF}}; since using {\cite[Eq. (1.110)]{Gradshteyn:2007}} we can write
  \begin{align}
 \Gamma^{F,\infty} = \left({{\prod\limits_{i = 1}^N {\left( {1 + {\kappa ^2}} \right) - 1} }}\right)^{-1} \approx \frac{1}{{N{\kappa ^2}}} = \frac{1}{N}\Gamma^{D,\infty}
 \end{align}
Similar to the AF relaying, we can obtain the diversity order of the HI system with the DF relaying.
{Utilizing {\cite[Eq. (1)]{Wang}} and using \eqref{N-hop-DF-HI2} when $\gamma > 1/{\delta ^2}$, it can be shown that when ${{\overline \Gamma  }_i}\gg1 \quad \forall i=1,...,N$, the diversity order for the DF relaying with the non-ID is given by }
\begin{align}
&{G_d^{\rm D}} = \min \Big\{{{{\beta'_{1} }}},2{{{\beta'_{2} }}},...,N{{{\beta'_{N} }}}\Big\}
 \label{Gd_DF}
\end{align}
which is the same as the diversity order of semi-blind AF relaying in \eqref{Gd_AF2}.
{{The diversity order in (20) and (37) can be reduced to some special cases reported in the literature (see for example \cite{jocn2022}-\cite{EGG},\cite{TWC2020}).}}

\subsection{Outage Probability of DF Relaying}
Using the CDF in \eqref{N-hop-DF-HI}, the OP of DF relaying can be readily obtained as
\begin{align}
{P_{out}^{\rm HI, D}}\left( {{\gamma _{th}}} \right) =Pr\left({\Gamma^D}< {{\gamma _{th}}} \right)= {F_{{\Gamma^D }}^{\rm HI}}\left( {{\gamma _{th}}} \right)
\end{align}
 \subsection{Average Bit Error Probability of DF Relaying}
{Again, due to the statistical dependence between numerator and denominator in {\eqref{sndr_df}}, the BEP for the HI cannot be obtained, analytically.}
Based on the expression in \eqref{N-hop-DF-HI-approx}, we can derive the average BEP of MH relaying with ID.
The BEP of $N$-hop DF relaying with the ID by substituting \eqref{N-hop-DF-HI-approx} in \eqref{Pe} and assuming $\kappa_i=0\quad \forall i=1,...,N$, is given by
\begin{align}
\overline P_e^{{\rm D, {\rm ID}}}=& \frac{\boldsymbol\delta}{2\Gamma \left( p \right)}\sum\limits_{k = 1}^{\mathbf{n}}\sum\limits_{i = 1}^N \sum_{l_i=1}^{\alpha_i}\frac{\rho _i}{\varrho _i}
\nonumber\\
\times &
{\bf{H}}_{{p_i} + 2,{q_i} + 1}^{{m_i},{n_i} + 2}\left[ {\frac{\varrho _i}{q_k} \left| {\begin{array}{*{20}{c}}
{\left( {[1-p,1,{a_i} ],[1,1,{A_i}]} \right)}\\
{\left( {[{b_i} ,0],[{B_i},1]} \right)}
\end{array}} \right.} \right]
\label{Pe-N-hop-DF}
\end{align}

\subsection{Ergodic Capacity of DF Relaying}
Similar to the AF relaying, the EC of the DF relaying is derived. By applying \eqref{sndr_df} to the PDF definition of the EC in Section II.D,
\begin{align}
\overline C^{\rm D, {\rm{HI}}} =& \underset{i\in \left\{ 1,...,N \right\}}{\mathop{\min }} \frac{1}{N} \mathbb{E}\Bigg\{ \log_2  \left(\frac{\Gamma _{i}}{{\kappa^2 _{i}}{\Gamma _{i}}+1}\right)  \Bigg\}\nonumber\\
\approx & \underset{i\in \left\{ 1,...,N \right\}}{\mathop{\min }}\, \frac{1}{N}\log_2 \left(\frac{\mathbb{E}\{{\Gamma} _{i}\}  }{{\kappa^2 _{i}}{\mathbb{E}\{{\Gamma} _{i}\}}+1}\right)
\label{EC_D_HI}
\end{align}
where we use the Jensen inequality in \eqref{EC_D_HI}. {By applying {\eqref{ceiling_DF}} to {\eqref{cap}}, we obtain SNDR capacity ceiling of DF relaying as
\begin{align}
{\overline C ^{D,\infty }} = \frac{1}{N}{\log _2}\left( {1 + c{\Gamma ^{D,\infty }}} \right).
\end{align}

\section{Optimization on The Level of Hardware Impairments}
In order to have low cost hardware, we formulate the optimization problem for determining the minimum level of HI at the $i^{\rm th}$ hop, represented by the decision variable $\kappa_i$, for AF and DF relaying. The cost of the hardware determines the level of hardware impairment such that the lower the cost, the higher the HI.
The objective of the optimization problem is to minimize the outage performance for a fixed cost of HI.
Note that an arbitrary SNDR analysis is intractable, therefore, we elaborate on the asymptotically low outage regime.

The optimization problem for AF and DF relaying is given as follows:
\begin{subequations}
\begin{align}
\min\limits_{\kappa_i, i \in [1,N]} &\hspace{3mm} {\rm{    }}{F_{{\Gamma ^Y}}^{\rm HI}}(\gamma)
\label{opt10}
\\
{\rm{s}}{\rm{.t}}{\rm{. }}
&\hspace{5mm}\sum\limits_{i = 1}^N {\kappa_i^2= A},
\label{opt11}
\\
&
\hspace{5mm}\kappa_i\geq0, \quad i \in [1,N].
\label{opt12}
\end{align}
\label{optB}
\end{subequations}
\hspace{-2mm}where $Y\in \{F,D\}$ and $A$ is a constant. Eq. \eqref{opt10} represents the OP as a function of the HI level of the nodes. Eq. \eqref{opt11} shows a total constraint on the HI level of all nodes, while Eq. \eqref{opt12} states the lower bound of the HI level.
{Since $\kappa_{i,t}$, $\kappa_{i,r}$ are the parameters that express the non-idealities of the transceivers, the decision variables in the optimization problem in \eqref{optB} are $\kappa_i$ $\forall i=1,...,N$.}
 Next, we solve this optimization problem for AF and DF relaying in Sections V.A and V.B, respectively.

\subsection{AF Relaying}
For the AF relaying, since the CDF in \eqref{CDF_HI_N_hsnr} is twice differentiable with respect to $\kappa_i\quad \forall i=1,...,N$ and due to convexity of the optimization problem in \eqref{optB}, we can use Karush-Kuhn-Tucker (KKT) conditions and obtain \eqref{sol_AF} $\forall j=1,...,N-1$
\begin{align}
&\sum\limits_{{l_1} = 1}^{{\alpha _1}} {\sum\limits_{{i_1} = 1}^{{m_1}} {{{D_{1,{i_1}}}}{{\left( {\frac{\gamma }{{{{\bar \Gamma }_1}}}} \right)}^{{\beta _{1,{i_1}}}}}} } \Bigg( \frac{{A'_N}}{{{{\left( {1 - {d_1}\gamma } \right)}^{{\beta _{1,{i_1}}}}}}}
\nonumber\\
&+
 \frac{{2\gamma {A_N}{\kappa _j}\left( {\kappa _N^2 - \kappa _j^2} \right)\left( {{d_1} + 1} \right)}}{{\left( {1 + \kappa _j^2} \right)\left( {1 + \kappa _N^2} \right){{\left( {1 - {d_1}\gamma } \right)}^{{\beta _{1,{i_1}}} + 1}}}} \Bigg) = 0,
\label{sol_AF}
\end{align}
where $A'_N = \frac{\partial A_N}{\partial \kappa_j}$ for $j=1,...,N-1$.
 However, no closed-form solution exists for $\bf H$-distribution. For the analysis of the Nakagami-$m$ fading channels obtained by setting ${\textit{\textbf{m}}_s}\rightarrow \infty, \boldsymbol\beta=\boldsymbol\beta_s=2$ (same as \cite{Alraddady}), the asymptotic CDF {\eqref{CDF_HI_N_hsnr}} can be approximated by keeping only the first term of {\eqref{cdf_H_hsnr0}}.
{Then, by applying the KKT conditions, we obtain}
\begin{subequations}
\begin{align}
\left(\kappa_1^{\rm opt}\right)^2 =&  {A + N - 1 - \left( {N - 1} \right)\sqrt[\mathlarger{\mathlarger{N}}]{{1 + \frac{1}{\gamma}}}},
\label{solution_AF0}
\\
\left(\kappa_i^{\rm opt}\right)^2 =&  {\sqrt[\mathlarger{\mathlarger{N}}]{{1 + \frac{1}{\gamma}}}-1} \quad \forall i=2,...,N
\label{solution_AF}
\end{align}
\end{subequations}
which is independent of the $i^{\rm th}$ link average SNR (${{\overline \Gamma  }_i} \quad \forall i=1,...,N$) and fading severities (i.e., ${\textit{\textbf{m}}}_i \quad \forall i=1,...,N$), while the OP depends on them. {This is reasonable since the $\kappa_i$ expression in the objective function does not depend on ${{\overline \Gamma  }_i} $ and ${\textit{\textbf{m}}}_i$.} The level of HI ($\kappa_i^{\rm opt} \quad \forall i=1,...,N$) is monotonically decreasing with respect to $N$.
 In addition, as $\gamma$ increases, $\kappa_j^{\rm opt} \quad \forall j=2,...,N$ decreases, whereas $\kappa_1^{\rm opt}$ increases. When $\gamma$ grows with no bound, the optimal solution is given by $\kappa_1^{\rm opt}=\sqrt{A}$, $\kappa_j^{\rm opt}=0 \quad \forall j=2,...,N$, i.e., we should use an ID in all hops except the first one.
From \eqref{solution_AF}, we can see that the equal level of HI is not always the optimal solution. For the equal level of HI, we should set the threshold SNDR as $\gamma=\frac{{{N^N}}}{{{{\left( {A + N} \right)}^N} - {N^N}}}$ .

\subsection{DF Relaying}
For the DF relaying employing \eqref{N-hop-DF-HI2}, again, the problem is convex. Therefore, by using the KKT conditions, we can obtain
 \begin{align}
&\sum_{l_i=1}^{\alpha_i}\sum\limits_{j = 1}^{{{{{m}}}_i}} \frac{{{D_{i,j}}} {\beta_{i,j}}\kappa _i}{ {\left( {{{\overline \Gamma  }_i}} \right)^{{\beta _{i,j}}}}}  {\left( {\frac{\gamma}{{1 - \kappa _i^2\gamma}}} \right)^{{\beta_{i,j}} + 1}}=\sum_{l_N=1}^{\alpha_N}\sum\limits_{j = 1}^{{{{{m}}}_N}} {\beta_{N,j}}\kappa _N
\nonumber\\
\times&\frac{{{D_{N,j}}}}{{\left( {{{\overline \Gamma  }_N}} \right)^{{\beta _{N,j}}}}}{\left( {\frac{\gamma}{{1 - \kappa _N^2\gamma}}} \right)^{{\beta_{N,j}} + 1}}{\rm{    }},\forall i = 1,...,N - 1
\end{align}
For the i.i.d. fading and symmetric channels where ${{{\alpha}}}_1=...={{{\alpha}}}_N={{{\alpha}}}$, ${{{m}}}_1=...={{{m}}}_N={{{m}}}$, $D_1=...=D_N=D$, $\beta_1=...=\beta_N=\beta$, and ${\overline \Gamma  _1} = {\overline \Gamma  _2} = ... = {\overline \Gamma  _N}$ we can obtain the optimal values of $\kappa _i$ which is
$\kappa _1^2 = \kappa _2^2{\rm{ = }}...{\rm{ = }}\kappa _N^2{\rm{ = }}\frac{{\rm{A}}}{N}{\rm{ }}$.
Of course, for the i.n.i.d. fading, this solution does not hold.\\
For Nakagami-$m$ fading, which is a special case of EGK fading with ${{\textit{\textbf{m}}}}_{s_i}\rightarrow \infty, \boldsymbol\beta_i={\boldsymbol\beta}_{s_i}=2, i \in [1,N]$  by assuming ${\textit{\textbf{m}}}_1={\textit{\textbf{m}}}_2=...={\textit{\textbf{m}}}_N={\textit{\textbf{m}}}$, we obtain
\begin{align}
\left(\kappa _i^{{\rm{opt}}}\right)^2 = {\frac{1}{\gamma} + \frac{{A\gamma - N}}{\gamma}\frac{{\prod\limits_{j = 1\hfill\atop
j \ne i\hfill}^N {\sqrt[{{\textit{\textbf{m}}} + 1}]{{\overline \Gamma _j^{\textit{\textbf{m}}}}}} }}{{\sum\limits_{j = 1}^N {\prod\limits_{k = 1\hfill\atop
k \ne j\hfill}^N {\sqrt[{{\textit{\textbf{m}}} + 1}]{{\overline \Gamma _k^{\textit{\textbf{m}}}}}} } }}}  \quad \forall i=1,...,N
\label{theta_DF_Gamma}
\end{align}
More hardware quality should be provided for the hops of lower SNR.
 Moreover, by increasing $N$, $\kappa_i^{\rm opt} \quad \forall i=1,...,N$ decreases, where after a certain $N$, the OP saturates.
 This means that from a certain point on, no matter how much we improve the quality of the hardware, it will have no effect on reducing the OP.
  Similar to AF relaying, as $\gamma$ increases, $\kappa_j^{\rm opt} \quad \forall j=2,...,N$ increases, whereas $\kappa_1^{\rm opt} $ decreases.
When $\gamma$ grows with no bound, $\left({\kappa_i^{\rm opt}}\right)^2=A\frac{{\prod\limits_{j = 1\hfill\atop
j \ne i\hfill}^N {\sqrt[{{\textit{\textbf{m}}} + 1}]{{\overline \Gamma _j^{\textit{\textbf{m}}}}}} }}{{\sum\limits_{j = 1}^N {\prod\limits_{k = 1\hfill\atop
k \ne j\hfill}^N {\sqrt[{{\textit{\textbf{m}}} + 1}]{{\overline \Gamma _k^{\textit{\textbf{m}}}}}} } }} \quad \forall i=1,...,N$. The equal optimum (i.e., $\left({\kappa_i^{\rm opt}}\right)^2=\frac{A}{N} \forall i=1,...,N$) only occurs when we have equal SNR in all links. On the other hand, when $i^{\rm th}$ link's SNR goes to infinity while other links' SNRs remain constant, $\left({\kappa_i^{\rm opt}}\right)^2=A+\frac{1-N}{\gamma} \quad \forall i=1,...,N$.
When ${{\overline \Gamma }_j}=\mu {{\overline \Gamma }_i} \quad \forall j \neq i$ with $\mu>1$ and ${{\overline \Gamma }_i}\rightarrow \infty$, $\left({\kappa_i^{\rm opt}}\right)^2=\frac{1}{\gamma} + \frac{{A\gamma - N}}{\gamma}\frac{{{\mu ^{\frac{{\textit{\textbf{m}}}}{{{\textit{\textbf{m}}} + 1}}}}}}{{{\mu ^{\frac{{\textit{\textbf{m}}}}{{{\textit{\textbf{m}}} + 1}}}} + N - 1}}$. This means that the lower the values of ${\overline \Gamma _i}$ are, the better the hardware quality should be. {To maintain the optimized HI level within the 3GPP limit, we can select proper values of $A$ and $\gamma$.}

\section{Performance Evaluation}
The goal of this section is to validate the accuracy of the analysis with Monte-Carlo (MC) simulations by numerically evaluating $\bf H$-functions and illustrate the concepts of SNDR and capacity ceilings, the necessary conditions, the guidelines for designing the MH hardware impaired systems and the optimization results.

{We use MATLAB, where the fading channel coefficient, and the distortion and receiver noises of all links are randomly generated.
Then, the received signal at the destination node is derived respectively for AF and DF relaying using {\eqref{E3_1}} and {\eqref{E4_1}}.
For all cases, $10^6$ realizations of the RVs are generated to perform the MC simulations.}
We consider respectively for weak, moderate, and strong turbulence conditions of each DGG+PE distributed FSO link  ${\Omega _1} = 1.0676,{\Omega _2} = 1.06,{\boldsymbol\alpha _1} = 2.1,{\boldsymbol\alpha _2} = 2.1,{\boldsymbol\beta _1} = 4,{\boldsymbol\beta _2} = 4.5$;  ${\Omega _1} = 1.5793,{\Omega _2} = 0.9671,{\boldsymbol\alpha _1} = 2.169,{\boldsymbol\alpha _2} = 1,{\boldsymbol\beta _1} = 0.55,{\boldsymbol\beta _2} = 2.35$; ${\Omega _1} = 1.5074,{\Omega _2} = 0.928,{\boldsymbol\alpha _1} = 1.8621,{\boldsymbol\alpha _2} = 1,{\boldsymbol\beta _1} = 0.5,{\boldsymbol\beta _2} = 1.8$; \cite{TWC2020}.
Unless otherwise stated, we assume equal average SNR for all links, ${\overline \Gamma }_{{{i}}}= {{\rm SNR}\hspace{1mm} \forall i\in \left\{ 1,...,N \right\}}$.

Fig. \ref{fig:Fig. V3} shows the E2E OP as a function of ${\rm SNR}$ for both ID and HI cases with the AF relay.
The OP in this figure is given in Eq. \eqref{N-hop-AF-HI}.
We have used $N=2$ FSO links for {both PE and no pointing errors (NPE)} and both HD and DD.
  For the HI, we assume $\kappa_1=\kappa_2=0.3$, $\gamma_{th}=2^2-1=4$. {According to \cite{ETSI1}, $3^{\rm rd}$ generation partnership project (3GPP) 5G new radio (5G NR) has impairment requirements in the range $\kappa_t \in [0.035, \hspace{1mm} 0.175]$ ($\kappa \in [0.0495, \hspace{1mm} 0.2475]$), where to achieve the highest values of the spectral efficiency, smaller values of HI are required.}
The HI degrades the OP {by 5 dB}, since the distortion noises of HI reduce the E2E SNDR of the multi-hop system.
Besides, the complex HD receiver performs better than DD {by 10 dB}. On the other hand, PE degrades the performance {by 7 dB in the medium SNRs and by 20 dB at high SNRs. }
The diversity orders confirm the derived expression of \eqref{Gd_AF}. Furthermore, the non-ID has the same slope (i.e. diversity order) as the ID; hence, HI causes only a shift of the OP curve to the right.
 \begin{figure}
\centering
\includegraphics[keepaspectratio,width=7cm]{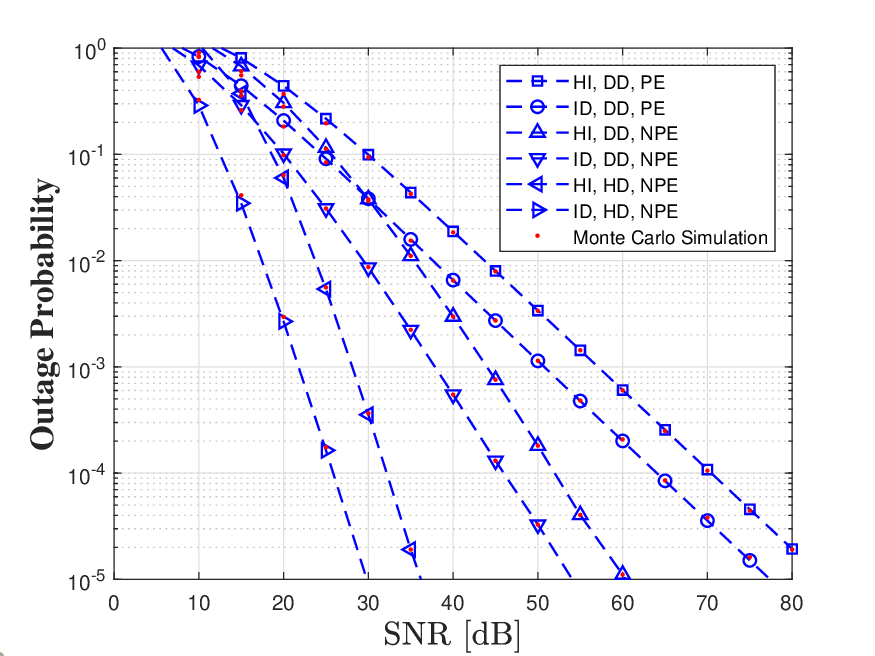}
\caption{OP versus ${\rm SNR}$ for both ID and HI.}
\label{fig:Fig. V3}
\end{figure}

Fig. \ref{fig:Fig. V4} shows the E2E OP as a function of SNDR threshold with AF and DF relays.
The OP in this figure is given in Eq. \eqref{N-hop-AF-HI}.
We consider ID and hardware with impairments of level of $\kappa_1=\kappa_2=0.15$ and $\kappa_1=\kappa_2=0.3$.
We have used the $N=3$ THz links with $\boldsymbol\alpha=1.4, \boldsymbol\mu=1.5$ \cite{Jiayi} and PE with $\xi=2.34$.
The OPs for low thresholds have not decreased much by HI.
However, there is an opposite action as the threshold SNR increases: the ID case slowly approaching to the full outage,
meanwhile the HI undergo a rapid approach toward the designated SNDR ceilings.
The values of these ceilings are consistent with our derived expressions in \eqref{ceiling_AF} for AF, i.e., $- 10\log \left( {\kappa _1^2\kappa _2^2\kappa _3^2 + \kappa _1^2\kappa _2^2 + \kappa _1^2\kappa _3^2 + \kappa _2^2\kappa _3^2 + \kappa _1^2 + \kappa _2^2 + \kappa _3^2} \right)=11.6$ and $5.3$ dB for $\kappa_i=0.15$ and $\kappa_i=0.3$, respectively.
For DF relaying, the values of these ceilings for $\kappa_i=0.15$ and $\kappa_i=0.3$ are equal to $ - 20\log \left( {\max \left( {\kappa _1^{},\kappa _2^{},\kappa _3^{}} \right)} \right)=16.5$ and $10.5$ dB, respectively.
\begin{figure}
\centering
\includegraphics[keepaspectratio,width=7cm]{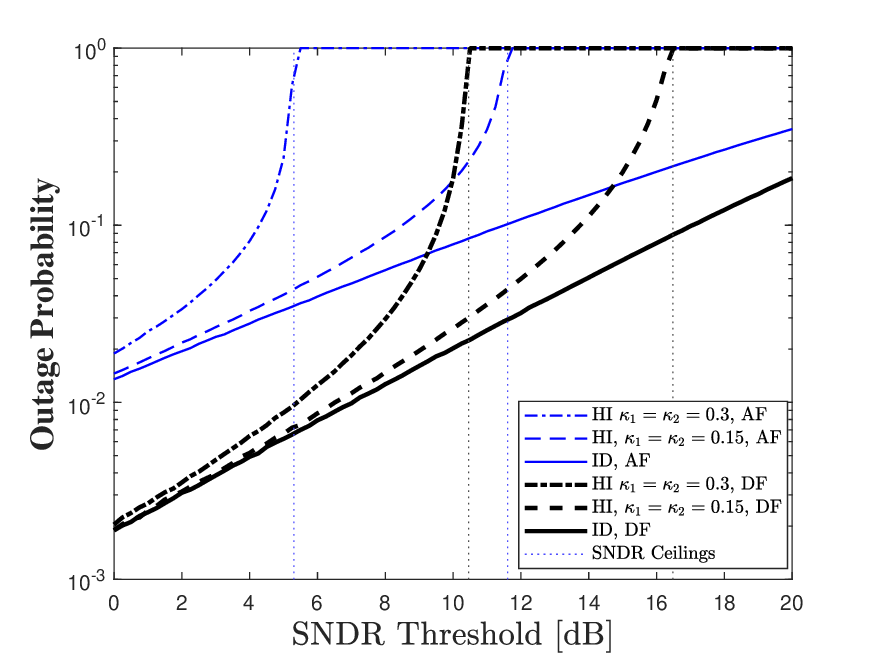}
\caption{OP versus $\gamma_{th}$ for both ID and HI.}
\label{fig:Fig. V4}
\end{figure}

Fig. \ref{fig:Fig. V5} shows the E2E OP of hardware-impaired dual-hop relaying as a function of $\kappa_1$ over moderate and weak $\mathcal{F}$ with PE for both semi-blind FG AF and DF relay. The OP in this figure is given in Eq. \eqref{N-hop-AF-HI} for the AF relaying and in Eq. \eqref{N-hop-DF-HI} for the DF relaying.
We assume different levels of impairments $\kappa_1$, $\kappa_2$ for which $\kappa_1+\kappa_2=1$. For moderate and weak $\mathcal{F}$ turbulence, we respectively consider ${\textit{\textbf a}} = 2.3378, {\textit{\textbf b}} = 4.5323$; and ${\textit{\textbf a}} = 4.5916, {\textit{\textbf b}} = 7.0941$  \cite{Badarneh}.
Despite the asymmetric SNRs (${\rm SNR}_1=30$ dB, ${\rm SNR}_2=20$ dB), the OP of AF case has been minimized by setting $\kappa_1 = \kappa_2 = 0.5$.
However, the OP of DF case has been minimized when the hop with the lower SNR has much more hardware quality.
Furthermore, when either the first hop or the second hop is ideal (i.e., $\kappa_1 = 0$ or $\kappa_2 = 0$, respectively), the system cannot work; therefore,
equipping the best quality hardware for one hop and neglecting the other one doesn't improve the OP at all.
\begin{figure}
\centering
\includegraphics[keepaspectratio,width=7cm]{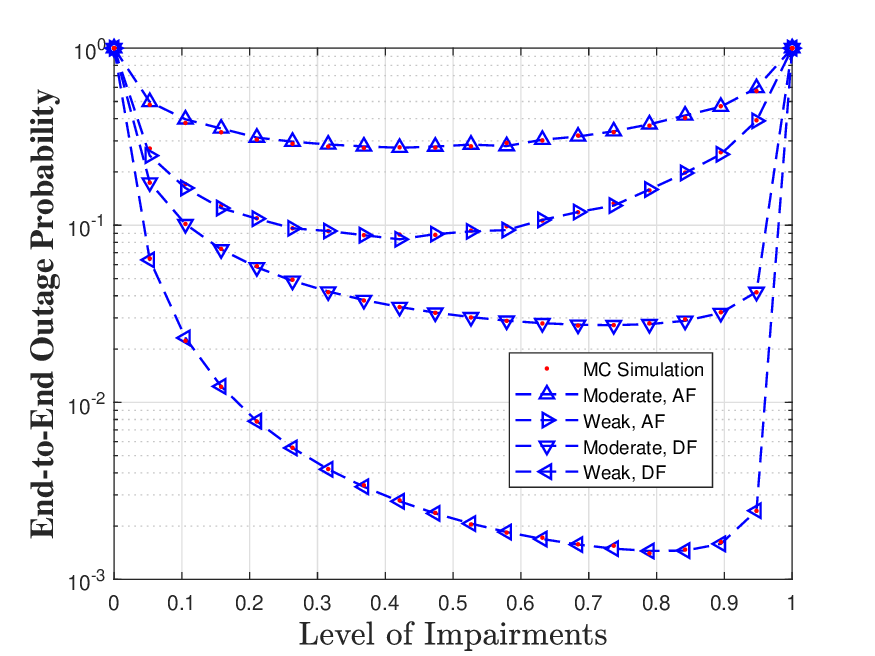}
\caption{OP versus asymmetric levels of impairments.}
\label{fig:Fig. V5}
\end{figure}

Fig. \ref{fig:Fig. V6} shows the E2E OP of dual-hop relaying as a function of $\kappa_i=\kappa \quad \forall i=1,...,N$ for both semi-blind FG AF and DF relay.
The OP in this figure is given in Eq. \eqref{N-hop-AF-HI} for the AF relaying and in Eq. \eqref{N-hop-DF-HI} for the DF relaying.
 We assume Nakagami-$m$ fading with ${\textit{\textbf{m}}}_i=2$ and the SNDR equal 15.
This figure shows the necessary but not sufficient conditions that behave as upper limits on the level of hardware defects when the OP is lower than one.
These necessary but not sufficient conditions determine the range in which the level of impairments must be.
For $N=2, 3 , 4$ and AF relaying, the HI levels are   $\kappa \le \sqrt{\sqrt{{1 + \frac{1}{x}}} - 1}=0.18$, $\kappa \le \sqrt{\sqrt[3]{{1 + \frac{1}{x}}} -1}=0.15$, $\kappa \le \sqrt{\sqrt[4]{{1 + \frac{1}{x}}} -1}=0.13$, respectively; while for the DF relaying, it is equal to $\kappa^2 \le \frac{1}{x} =0.26$.
Focusing on the four lower curves and requiring that $P_{out}(15) \leq 10^{-1}$, we can identify two possible hardware operating regimes: 1) FG AF relaying with $\kappa \leq 0.14$, for $N=2$; 2) DF relaying with $\kappa \leq 0.24, 0.23, 0.22$ respectively for $N=2,3,4$. The different acceptable levels of impairments show that DF relaying is more robust to HI comparing with AF relaying, hence, can work with lower quality hardware. By using multiple hops, we gain more coverage.
\begin{figure}
\centering
\includegraphics[keepaspectratio,width=7cm]{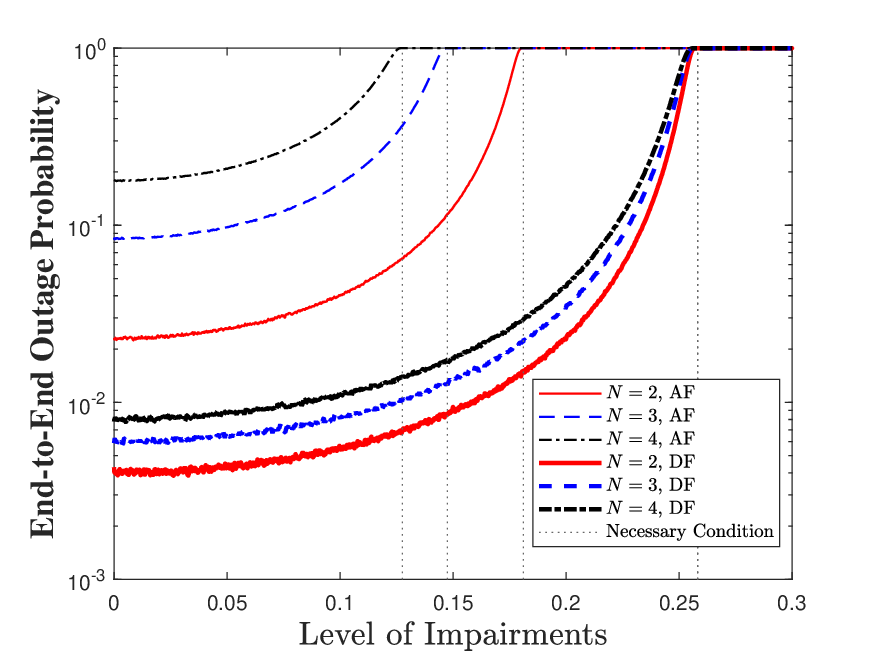}
\caption{OP versus symmetric levels of hardware impairments.}
\label{fig:Fig. V6}
\end{figure}

Fig. \ref{fig:Fig. V7} shows the E2E BEP of triple-hop AF relaying as a function of ${\rm SNR}$ over different combinations of fading (i.e., RF, FSO, and THz). {We have used the on-off keying (OOK) modulation which works in both radio wireless and optical wireless systems.}
The BEP in this figure is given in Eq. \eqref{Pe_HVLC}. We assume 7 different cases of FSO-FSO-FSO, FSO-RF-FSO, THz-FSO-THz, RF-FSO-RF, RF-THz-FSO, RF-RF-FSO, and RF-RF-RF. For the RF links, we have shadowed EGK with ${\textit{\textbf{m}}}={\textit{\textbf{m}}}_s=\boldsymbol\beta=\boldsymbol\beta_s=2$. For the FSO links, we consider $\xi=1.22$, $r=2$, and strong DGG turbulence with ${\boldsymbol\alpha _1} = 1.8621,{\boldsymbol\alpha _2} = 1,{\boldsymbol\beta _1} = 0.5,{\boldsymbol\beta _2} = 1.8,{\Omega _1} = 1.5074,{\Omega _2} = 0.928$. For the THz links, $\boldsymbol\alpha=1.4, \boldsymbol\mu=1.5$, and $\xi=1.22$.
FSO-FSO-FSO combination has the worst performance, while RF-RF-RF combination has the best performance,
{since the FSO links are under strong turbulence, while RF links are subject to weak fading channels. }
\begin{figure}
\centering
\includegraphics[keepaspectratio,width=7cm]{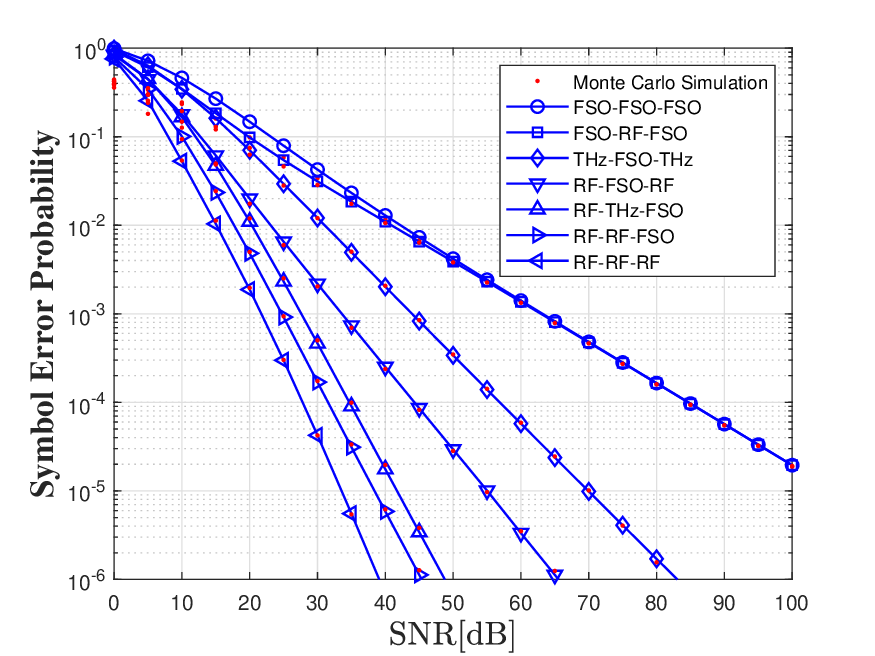}
\caption{BEP of triple-hop AF relaying versus ${\rm SNR}$.}
\label{fig:Fig. V7}
\end{figure}

 Fig. \ref{fig:Fig. V8} shows the approximated OP of $N$-hop AF relaying as a function of ${\rm SNR}$ for different values of $N$ and the threshold SNDR.
The OP in this figure is given in Eq. \eqref{CDF_HI_N_hsnr}.
We assume dual-hop ($N=2$) and triple-hop ($N=3$) relaying with $x=1, x=15$ and $\kappa_i = 0.1 \quad \forall i=1,...,N$.
The approximated OP expression in \eqref{CDF_HI_N_hsnr} predicts well the simulation results, with perfect match at high SNRs.
 As expected, by increasing the threshold SNDR $x$, the performance degrades. In addition, if we put more relays between source and destination, the performance improves. In fact, using multiple relays reduces the OP by scaling the fading.
\begin{figure}
\centering
\includegraphics[keepaspectratio,width=7cm]{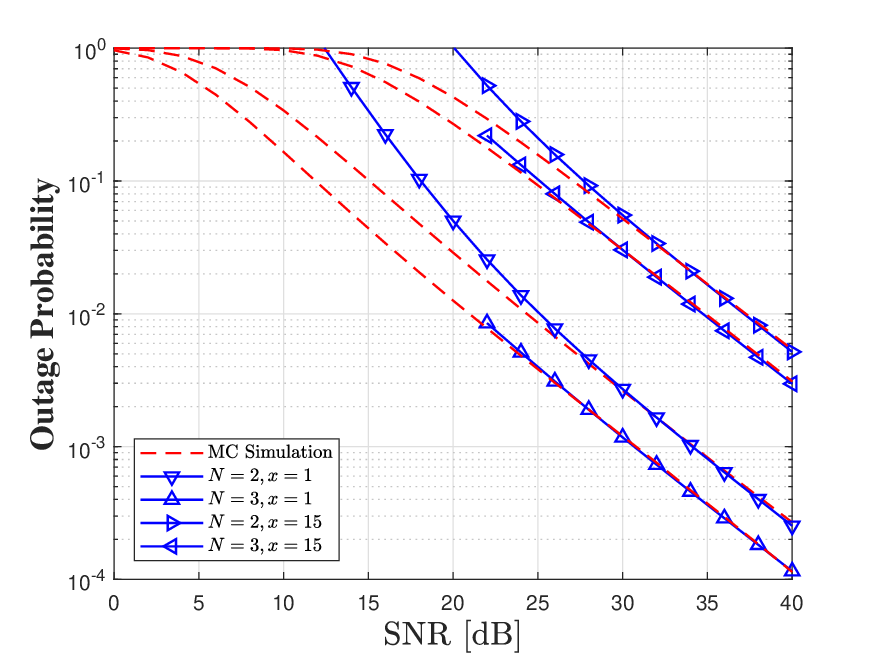}
\caption{OP of AF relaying versus ${\rm SNR}$ for different values of $N$ and $x$.}
\label{fig:Fig. V8}
\end{figure}

Fig. \ref{fig:Fig. V9} shows the E2E OP as a function of ${\rm SNR}$ for $N$ FSO with DD receiver and AF relay over Malaga turbulence channels {for different values of PE and different relay locations.}
The OP in this figure is given in Eq. \eqref{N-hop-AF-HI}.
We assume a 3000m distance between the source and the destination with dual-hop ($N=2$) and triple-hop ($N=3$) relaying with $\xi=1.22, 7.35$. For $N=2$, the link lengths are $L$=[2500 500], $L$=[1500 1500], and $L$=[500 2500]. For $N=3$, the link lengths are $L$=[1000 1000 1000], and $L$=[500 1000 1500]. The closer the first relay to the source, the better the performance.
\begin{figure}
\centering
\includegraphics[keepaspectratio,width=7cm]{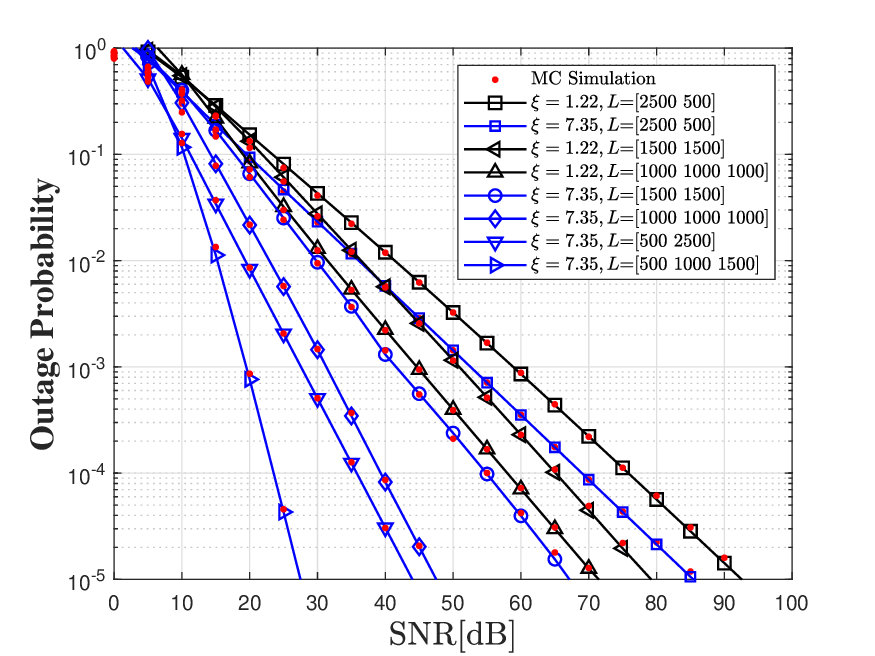}
\caption{OP versus ${\rm SNR}$ for different relay locations and $\xi$.}
\label{fig:Fig. V9}
\end{figure}

Fig. \ref{fig:Fig. V10} shows the EC of dual-hop and triple-hop relaying as a function of ${\rm SNR}$ over multiple FSO links with 3000m distance between the source and the destination for both AF and DF relay.
The EC in this figure is given in Eq. \eqref{EC_F_ID} for the AF relaying with ID, in Eq. \eqref{EC_F_HI} for the AF relaying with HI, and in Eq. \eqref{EC_D_HI} for the DF relaying with HI and ID. For hardware with impairments, we assume $\kappa_i=0.1 \quad \forall i=1,...,N$.
We can see a perfect match between simulation and analytical results at medium and high SNRs which indicates the accuracy of the derived analytical expressions.
 As observed, the EC of HI
saturates and approaches $\frac{1}{N}{\log _2}\left( {1 + {{\left( {\prod\limits_{i = 1}^N {\left( {1 + {\kappa _i}} \right) - 1} } \right)}^{ - 1}}} \right)$ and $\frac{1}{N}{\log _2}\left( {1 + \min \left( {{\kappa _1},...,{\kappa _N}} \right)} \right)$ respectively for AF and DF relaying, as we proved in Section III.D and Section IV.D.
As the asymptotic EC is assessed by the HI level, it increases when $\kappa_i$ for $i=1,...,N$ decreases.
For the ID, as the SNR increases, the performance gets better with no bound.
 As expected, HI's influence at low SNRs is small in contrast with high SNRs. More importantly, increasing the number of relays from 1 to 2 leads to a performance degradation. More precisely, at ${\rm SNR}$=40 dB, the EC of HI is equal to 2.8 Bits/Sec/Hz for dual-hop AF relaying, while this reduces to 1.7 Bits/Sec/Hz for triple-hop AF relaying. This is because the performance degradation due to the imperfection of hardware in triple-hop case dominates the advantage of reducing the links' length. Similar to the OP, the EC performance of HI with DF relaying is better than the HI with the AF relaying.
\begin{figure}
\centering
\includegraphics[keepaspectratio,width=7cm]{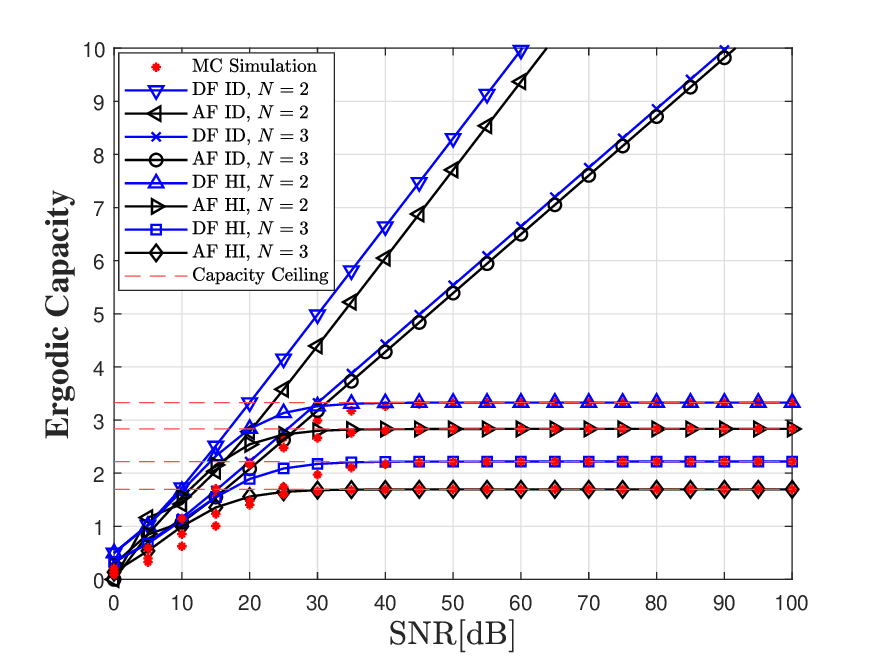}
\caption{EC versus ${\rm SNR}$ for the ID and HI cases.}
\label{fig:Fig. V10}
\end{figure}

{Fig. \ref{fig:Fig. V11} shows the OP of multi-hop DF relaying versus SNR for $N=2,3,4$. We consider both the optimal solution as in \eqref{theta_DF_Gamma} and equal $\kappa_i$ for Nakagami-$m$ fading channels. We assume $\Gamma_1 = {\rm SNR}$, $\Gamma_i=5\Gamma_1$ $\forall i>1$. Other parameters are $A=0.3$, ${\textit{\textbf{m}}}=2$, and $ \gamma  = 2^{N r}-1$ with $r=1$ (i.e., 1 bit/channel use).
As observed, the optimal solution has more than one dB performance gain than the equal solution which verifies the supremity of the optimization problem on $\kappa_i$.
This is reasonable since the optimal solution needs to satisfy the demand of all the links according to their SNR.
For example, at ${\rm SNR}$=20 dB, the optimal OP of $N=2$ is 0.003, while the non-optimal OP is 0.0055. The optimal OP of $N=3$ is 0.05, while the non-optimal OP is 0.09.
Also, it can be seen that the relative distance between the curves increases for higher values of the $N$. This implies that the impact of $N$ becomes increasingly more pronounced.}
\begin{figure}
\centering
\includegraphics[keepaspectratio,width=7cm]{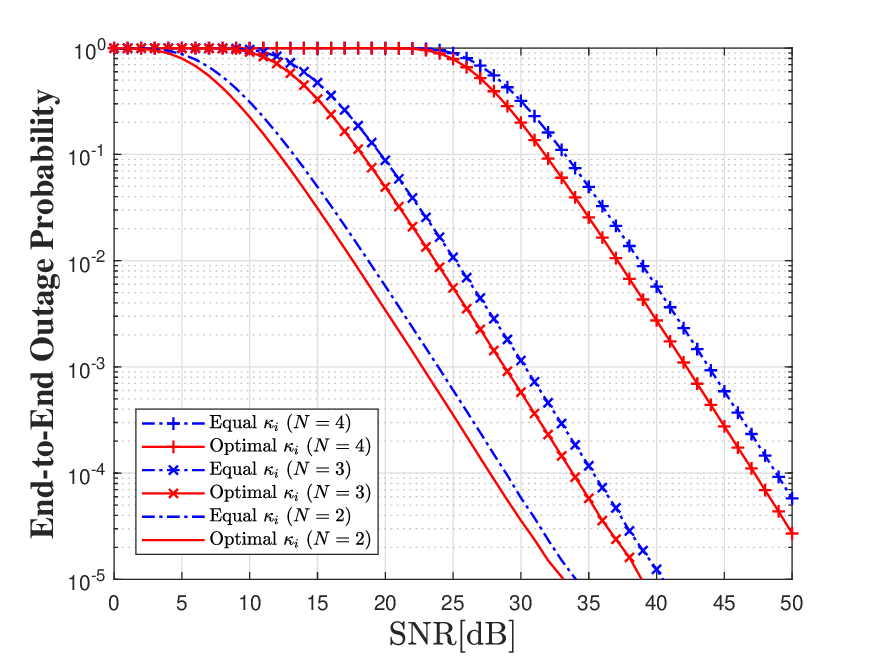}
\caption{{OP versus ${\rm SNR}$ with the optimal and equal $\kappa_i$.}}
\label{fig:Fig. V11}
\end{figure}

\section{Conclusion}
 We investigate the performance of multi-hop communications for both hardware imperfection and ID cases. By assuming the $\bf H$-fading model, which includes many fading models (e.g., RF, FSO, THz and MMW models), we derive closed-form expressions of OP, BEP, and EC in terms of single-variate and bivariate $\bf H$-functions. We also pursue an asymptotic analysis and derive the diversity order of system for different relaying protocols.
The derived expressions accurately describe the effects of HI and show that substantive ceilings of SNDR and capacity exists. These ceilings cannot be eliminated by altering the fading conditions or using more signal power. The SNDR threshold for the DF relaying is almost $N$ times the SNDR of FG AF relaying, when we assume equal level of HI in all nodes. Moreover, DF relaying is more robust to the HI at the cost of higher complexity compared to the AF relaying.
 Furthermore, SNDR ceiling exists in the high SNR regions and the value of this ceiling has an inverse relationship with the level of HI. This finding shows that hardware flaws intimately constrain both AF and DF relaying systems. Likewise, there is a capacity ceiling on the top of the EC.
 The results demonstrate that the optimal level of HI depends on the number of links and the threshold SNDR for AF relaying while it is independent of links' SNRs and fading parameters. For the DF relaying, the optimal level of HI depends on all links' SNRs and fading severity in addition to the number of links and the threshold SNDR.
In the future, we plan to work on the analysis of the HI systems with full-duplex relaying, while incorporating imperfect channels and imposing co-channel interferers on the relay nodes.

\vspace{-2mm}
\section*{APPENDIX I: Proof of Theorem 1}
By applying the average operator in \eqref{SNDR_AFF}, we obtain
\begin{align}
{\Gamma ^F} = \frac{{\prod\limits_{j = 1}^N {G_{j - 1}^2h_j^{{r_j}}{P_1}} }}{{\sum\limits_{j = 1}^N \left({\sigma _j^2}+ \sum\limits_{j = 1}^N {\mathbb{E}\left\{ {\eta _j^2} \right\}h_j^{{r_j}}} \right)\prod\limits_{k = j + 1}^N {G_{k - 1}^2h_k^{{r_k}}} }}
\label{V1}
\end{align}
From ${s_{i + 1}} = {G_i}{y_i} $, we have ${P_{i + 1}} = G_i^2\mathbb{E}\left\{ {y_i^2} \right\}$. Knowing $\eta_i \sim \mathcal{CN}(0; \kappa_i^2 P_i)$, we obtain
 $\mathbb{E}\left\{ {y_i^2} \right\}$.
Since ${y_i} = {\left( {{h_i}} \right)^{\frac{{{r_i}}}{2}}}{s_i} + \left( {{h_i}} \right)^{\frac{{{r_i}}}{2}}{\eta _i} + {v_i}$, we can write
\begin{align}
\mathbb{E}\left\{ {y_i^2} \right\} =&\hspace{2mm} G_{i - 1}^2h_i^{{r_i}}\left( {1 + \kappa _i^2} \right)\mathbb{E}\left\{ {y_{i - 1}^2} \right\} + \sigma _i^2 \nonumber\\
=&\hspace{2mm} {P_1}\prod\limits_{k = 1}^i {G_{k - 1}^2h_k^{{r_k}}\left( {1 + \kappa _k^2} \right)}
\nonumber\\
 + &\hspace{2mm}\sum\limits_{j = 2}^{i + 1} {\sigma _j^2\prod\limits_{k = j}^i {\left[ {G_{k - 1}^2h_k^{{r_k}}\left( {1 + \kappa _k^2} \right)} \right]} }
\end{align}
This leads to variance of the aggregated distortion noise for the $i^{\rm th}$ hop as follows:\hspace{3cm}
\begin{align}
&\mathbb{E}\left\{ {\eta _{i + 1}^2} \right\}
= \kappa _{i + 1}^2{P_1} \prod\limits_{k = 1}^i {\left[G_k^2 {h_k^{{r_k}}\left( {1 + \kappa _k^2} \right)} \right]}
   \nonumber\\
   &+ \kappa _{i + 1}^2\sum\limits_{l = 2}^{i + 1} {\sigma _{l-1}^2\prod\limits_{k = l}^{i + 1} {\left[ {G_{k - 1}^2} \right]} } \prod\limits_{k = l}^i {\left[ {h_k^{{r_k}}\left( {1 + \kappa _k^2} \right)} \right]}
\label{eta}
\end{align}
The $i^{\rm th}$ relay's amplification gain of the FG AF relaying can be written as
\begin{align}
  G_i^2 = \frac{{{P_{i + 1}}}}{{{P_i}\mathbb{E}\left\{ {h_i^{{r_i}}} \right\}\left( {1 + \kappa _i^2} \right) + \sigma _i^2}} = \frac{{{P_{i + 1}}/\sigma _i^2}}{{\left( {1 + \kappa _i^2} \right)\mathbb{E}\left\{ {{\Gamma _i}} \right\} + 1}}
\label{V2}
\end{align}
Defining $  {C_{{R_i}}} \triangleq \left( {1 + \kappa _i^2} \right)\mathbb{E}\left\{ {{\Gamma _i}} \right\} + 1$, we can obtain the gain as
$G_i^2 = \frac{{{P_{i + 1}}}}{{\sigma _i^2{C_{{R_i}}}}}$.\\
By multiplying the nominator and denominator of \eqref{V1} with $\frac{{\prod\limits_{k = 2}^N {{P_k}} }}{{\prod\limits_{k = 1}^N {\left[ {G_{k - 1}^2\sigma _k^2} \right]} }}$, and putting \eqref{eta} and \eqref{V2} into \eqref{V1}, we obtain
\vspace{-3mm}
\begin{align}
{\Gamma ^F} = \frac{{\prod\limits_{j = 1}^N {\left[ {{\Gamma _j}} \right]} }}{\begin{gathered}\Bigg[
  \sum\limits_{j = 1}^N {\kappa _j^2\prod\limits_{k = 1}^{j - 1} { {\left( {1 + \kappa _k^2} \right)} } } \prod\limits_{k = 1}^N { {{\Gamma _k}} }
   +
  \sum\limits_{j = 1}^N {\prod\limits_{k = 1}^{j - 1} { {{C_{{R_k}}}} } } \prod\limits_{k = j + 1}^N {{\Gamma _k}}
   \hfill \\
   + \sum\limits_{j = 2}^N {\kappa _j^2\sum\limits_{l = 2}^j {\prod\limits_{k = 1}^{l - 2} { {{C_{{R_k}}}} } } } \prod\limits_{k = l}^{j - 1} { {\left( {1 + \kappa _k^2} \right)}} \prod\limits_{k = l}^N {{{\Gamma _k}} } \Bigg]
\end{gathered} }
\end{align}
Finally, after some simple algebraic manipulations, we have
\begin{align}
{\Gamma ^F} = \left[{{\prod\limits_{j = 1}^N {\left[ {{\Gamma _j}} \right]} }}\right]/\left[{
  \prod\limits_{j = 1}^N { {{\Gamma' _j}} }    
   + \sum\limits_{j = 1}^N {\prod\limits_{k = 1}^{j - 1} { {{C_{{R_k}}}} } } \prod\limits_{k = j + 1}^N {{\Gamma' _k}}
}\right]
\end{align}
which is \eqref{SNDR_N}.

\section*{APPENDIX II: Proof of Theorem 2}
The closed-form CDF expression of the dual-hop FG AF relaying with the imperfection of hardware is provided to evaluate the OP.
By doing so, we are able to obtain the CDF of MH relaying using an inductive argument.

The E2E instantaneous SNDR of dual-hop relaying, by putting $N=2$ in \eqref{SNDR_N}, is obtained as
\begin{align}
\Gamma^{\rm HI, dual}_{F}=\frac{\Gamma_1 \Gamma_2}{d_1 \Gamma_1 \Gamma_2 + \lambda_2\Gamma_2+C_{R_1}},
\label{e2e_SNDR}
\end{align}
where $d_1 =\lambda_1-1=\kappa _1^2 + \kappa _2^2 + \kappa _1^2\kappa _2^2$, $\lambda_2=1+\kappa _2^2$ and $C_{R_1} \triangleq\mathbb{E}\left[\Gamma_1\right](1+\kappa_1^2)+1$.
\\
By substituting \eqref{e2e_SNDR} into the OP definition, we have
\begin{align}
F_{{\Gamma ^F}}^{{\rm{HI,dual}}}\left(\gamma\right)=Pr \left(\frac{\Gamma_1 \Gamma_2}{d_1 \Gamma_1 \Gamma_2 + \lambda_2\Gamma_2+C_{R_1}}\le \gamma\right),
\label{outage_Pr_HI}
\end{align}
After some straightforward mathematical manipulations, we obtain
\begin{align}
F_{{\Gamma ^F}}^{{\rm{HI,dual}}}\left(\gamma\right)=
{Pr \left({\Gamma_1 }\le \frac{\lambda_2\Gamma_2+C_{R_1}}{(1-d_1 \gamma)\Gamma_{2}}\gamma \right), \hspace{2mm} \gamma < 1/d_1}
\label{outage_Pr_HI2}
\end{align}
As anticipated, when $\gamma \ge 1/d_1$, the CDF simplifies to one.
Therefore, we assume the CDF threshold is strictly inferior to $1/d_1$ (i.e., $\gamma < 1/d_1$).\\
{
Because of independency of $\Gamma_1$ and $\Gamma_2$, the outage expression in \eqref{outage_Pr_HI2} can be written as follows:
\begin{align}
F_{{\Gamma ^F}}^{{\rm{HI,dual}}}\left(\gamma\right)=&\int_{0}^{\infty} \int_{0}^{\frac{\lambda_2}{(1-d_1 \gamma)}\gamma} f_{\Gamma_1} \left(y \right) f_{\Gamma_2}\left( x \right) dy dx
\nonumber \\
+&\int_{0}^{\infty} \int_{{\frac{\lambda_2}{(1-d_1 \gamma)}\gamma} }^{\frac{\lambda_2}{(1-d_1 \gamma)}\gamma+\frac{C_{R_1}}{(1-d_1 \gamma)x}\gamma} f_{\Gamma_1} \left(y \right) f_{\Gamma_2}\left( x \right) dy dx.
\label{outage_Pr_HI222}
\end{align}
By interchanging the order of integral, the outage in \eqref{outage_Pr_HI222} can be obtained as
\begin{align}
F_{{\Gamma ^F}}^{{\rm{HI,dual}}}\left(\gamma\right)=& F_{\Gamma_1} \left({\frac{\lambda_2}{(1-d_1 \gamma)}\gamma} \right)
\nonumber \\
 +&\int_{\frac{\lambda_2}{(1-d_1 \gamma)}\gamma}^{\infty} f_{\Gamma_1} \left(y \right) F_{\Gamma_2}\left( {\frac{\frac{C_{R_1}}{(1-d_1 \gamma)}\gamma}{y-\frac{\lambda_2}{(1-d_1 \gamma)}\gamma}} \right) dy.
\label{}
\end{align}
After the change of integral variable $y-\frac{\lambda_2}{(1-d_1 \gamma)}\gamma \to y$, we have
\begin{align}
F_{{\Gamma ^F}}^{{\rm{HI,dual}}}\left(\gamma\right) =& 1-\int_{0}^{\infty} f_{\Gamma_1} \left(y+\frac{\lambda_2}{(1-d_1 \gamma)}\gamma \right)
\nonumber \\
\times&
\left(1-F_{\Gamma_2}\left( {\frac{C_{R_1}}{(1-d_1 \gamma)y}\gamma} \right) \right)dy.
\label{outage_Pr_HI21}
\end{align}}
Due to independency of both link's SNR distributions, after some algebraic manipulations, the CDF in \eqref{outage_Pr_HI21} can be expressed as
\begin{align}\label{}
F_{{\Gamma ^F}}^{{\rm{HI,dual}}}\left(\gamma\right)=& F_{\Gamma_{1}} \left({\frac{\lambda_2\gamma}{1-d_1 \gamma}} \right)
 +\int_{0}^{\infty} f_{\Gamma_{1}} \left(y+\frac{\lambda_2\gamma}{1-d_1 \gamma} \right)
 \nonumber\\ \times &
F_{\Gamma_{2}}\left( {\frac{C_{R_1}\gamma}{(1-d_1 \gamma)y}} \right)dy
\label{int_f}
\end{align}
By substituting \eqref{pdf_H} and \eqref{cdf_H} into \eqref{int_f}, then expanding $\bf H$-functions as in \cite[Eq. (9.301)]{Gradshteyn:2007} to solve the integral in \eqref{int_f} we end up with \eqref{int555},
\begin{figure*}
\vspace{-4mm}
\begin{align}
{I_1} =& \frac{{ - 1}}{{4{\pi ^2}}}\int\limits_0^\infty  {\int\limits_{{\mathbb{C}_1}} {\int\limits_{{\mathbb{C}_2}} {\frac{1}{{\left(y+\frac{{{\lambda _2\gamma}}}{{1 - d_1\gamma }} \right)}}} } } \frac{{\prod\limits_{i = 1}^2 {\prod\limits_{k = 1}^{{m_i}} {\Gamma \left( {b_{i,k} - {B_{i,k}}{s_i}} \right)\prod\limits_{k = 1}^{{n_i}} {\Gamma \left( {1 - a_{i,k} + {A_{i,k}}{s_i}} \right)} } } }}{{\prod\limits_{i = 1}^2 {\prod\limits_{k = {m_i} + 1}^{{q_i}} {\Gamma \left( {1 - b_{i,k} + {B_{i,k}}{s_i}} \right)\prod\limits_{k = {n_i} + 1}^{{p_i}} {\Gamma \left( {b_{i,k} - {A_{i,k}}{s_i}} \right)} } } }}
\nonumber\\ \times &
\frac{{\Gamma \left( { - {s_2}} \right)}}{{\Gamma \left( {1 - {s_2}} \right)}}{\left[ {\frac{\varrho _1}{{{{\overline \Gamma  }_1}}}\left(y+\frac{{{\lambda _2\gamma}}}{{1 - d_1\gamma }} \right)} \right]^{{s_1}}}{\left[ {\frac{{{\varrho _2} }}{{{\overline \Gamma  }_2}y}\frac{{{C_{{R_1}}}\gamma}}{{1 - d_1\gamma }}} \right]^{{s_2}}}d{s_2}d{s_1}dy
\label{int555}
\end{align}
\hrulefill
\vspace{-4mm}
\end{figure*}
where $\mathbb{C}_1$ and $\mathbb{C}_2$ are the $s_1$-plane and the $s_2$-plane contours, respectively.
{Using the change of integral orders, we obtain \eqref{int5555} given in the next page.}
\begin{figure*}
\vspace{-3mm}
\begin{align}
&{
{I_1} = \frac{{ - 1}}{{4{\pi ^2}}}}  {\int\limits_{{\mathbb{C}_1}} {\int\limits_{{\mathbb{C}_2}} {} } } \frac{{\prod\limits_{i = 1}^2 {\prod\limits_{k = 1}^{{m_i}} {\Gamma \left( {b_{i,k} - {B_{i,k}}{s_i}} \right)\prod\limits_{k = 1}^{{n_i}} {\Gamma \left( {1 - a_{i,k} + {A_{i,k}}{s_i}} \right)} } } }}{{\prod\limits_{i = 1}^2 {\prod\limits_{k = {m_i} + 1}^{{q_i}} {\Gamma \left( {1 - b_{i,k} + {B_{i,k}}{s_i}} \right)\prod\limits_{k = {n_i} + 1}^{{p_i}} {\Gamma \left( {b_{i,k} - {A_{i,k}}{s_i}} \right)} } } }}
\nonumber\\ & {\times
\frac{{\Gamma \left( { - {s_2}} \right)}}{{\Gamma \left( {1 - {s_2}} \right)}}{\left[ {\frac{\varrho _1}{{{{\overline \Gamma  }_1}}}} \right]^{{s_1}}}{\left[ {\frac{{{\varrho _2} }}{{{\overline \Gamma  }_2}}\frac{{{C_{{R_1}}}\gamma}}{{1 - d_1\gamma }}} \right]^{{s_2}}}
\int\limits_0^\infty \frac{ {{\left(y+\frac{{{\lambda _2\gamma}}}{{1 - d_1\gamma }} \right)}}^{s_1-1} }{y^{s_2}}   dy
d{s_2}d{s_1}}
\label{int5555}
\end{align}
\hrulefill
\vspace{-4mm}
\end{figure*}
We can solve the integral on $y$ in \eqref{int5555} by using \cite[Eq. (3.194.3)]{Gradshteyn:2007}. The result can be written in terms of Gamma function by using \cite[Eq. (8.384.1)]{Gradshteyn:2007} given in \eqref{int5_2}.
\begin{figure*}
\vspace{-3mm}
\begin{align}
{I_1} =& \frac{{ - 1}}{{4{\pi ^2}}}\int\limits_{{\mathbb{C}_1}} {\int\limits_{{\mathbb{C}_2}} {\frac{{\prod\limits_{i = 1}^2 {\prod\limits_{k = 1}^{{m_i}} {\Gamma \left( {b_{i,k} - {B_{i,k}}{s_i}} \right)\prod\limits_{k = 1}^{{n_i}} {\Gamma \left( {1 - a_{i,k} + {A_{i,k}}{s_i}} \right)} } } }}{{\prod\limits_{i = 1}^2 {\prod\limits_{k = {m_i} + 1}^{{q_i}} {\Gamma \left( {1 - b_{i,k} + {B_{i,k}}{s_i}} \right)\prod\limits_{k = {n_i} + 1}^{{p_i}} {\Gamma \left( {b_{i,k} - {A_{i,k}}{s_i}} \right)} } } }}} }
\nonumber\\ \times &
\frac{{\Gamma \left( { - {s_2}} \right)\Gamma \left( {{s_2} - {s_1}} \right)}}{{\Gamma \left( {1 - {s_2}} \right)}}\frac{{\Gamma \left( {1 - {s_2}} \right)}}{{\Gamma \left( {1 - {s_1}} \right)}}{\left[ {\frac{{{\varrho _1}}}{{{{\bar \Gamma }_1}}}\frac{{{\lambda _2}\gamma }}{{1 - {d_1}\gamma }}} \right]^{{s_1}}}{\left[ {\frac{{{\varrho _2}}}{{{{\bar \Gamma }_2}}}\frac{{{C_{{R_1}}}}}{{{\lambda _2}}}} \right]^{{s_2}}}d{s_2}d{s_1}
\label{int5_2}
\end{align}
\hrulefill
\vspace{-4mm}
\end{figure*}
By change of integral variable $s_1 \rightarrow -s_1$ and using \cite[Eq. (2.57)]{mathai}, we obtain \eqref{dual-hop-HI-exact}.
\newline
{ By expanding of the Mellin-Barnes integrals involved in the bivariate H-function, the approximation can be obtained by calculating the residue of the corresponding integrands at the nearest pole to the contour.} Using the Cauchy's residue theorem, \eqref{int5_2} can be approximated as \eqref{OP_HI}.
\begin{figure*}
\vspace{-3mm}
\begin{align}
{I_1} {{\approx}} \frac{1}{{2\pi j}}\int\limits_{{\mathbb{C}_1}} {\frac{{\prod\limits_{i = 1}^2 {\prod\limits_{k = 1}^{{m_i}} {\Gamma \left( {b_{i,k} - {B_{i,k}}{s_1}} \right)\prod\limits_{k = 1}^{{n_i}} {\Gamma \left( {1 - a_{i,k} + {A_{i,k}}{s_1}} \right)} } } }}{{\prod\limits_{i = 1}^2 {\prod\limits_{k = {m_i} + 1}^{{q_i}} {\Gamma \left( {1 - b_{i,k} + {B_{i,k}}{s_1}} \right)\prod\limits_{k = {n_i} + 1}^{{p_i}} {\Gamma \left( {b_{i,k} - {A_{i,k}}{s_1}} \right)} } } }}} \frac{{\Gamma \left( { - {s_1}} \right)}}{{\Gamma \left( {1 - {s_1}} \right)}}{\left[ {\frac{{\varrho _1}{\varrho _2}}{{{{\overline \Gamma  }_1}}{{{\overline \Gamma  }_2}}}\frac{{{C_{{R_1}}}\gamma}}{{1 - d_1\gamma }} } \right]^{{s_1}}}d{s_1}
\label{OP_HI}
\end{align}
\hrulefill
\vspace{-4mm}
\end{figure*}
Finally, by applying \cite[Eqs. (1.1.1, 1.1.2)]{Kilbas} to \eqref{OP_HI} and using \eqref{int_f}, we obtain \eqref{2-hop-AF-HI}.

\section*{APPENDIX III: Proof of Theorem 3}
The closed-form CDF expression of hardware impaired systems is provided to evaluate the OP, BEP, and EC of MH relaying networks.
For a dual-hop relaying, we can write
\begin{align}
\Gamma _{N - 1,N}^F = \left({{{d_{N - 1}} + \frac{{{\lambda _N}}}{{{\Gamma _{N - 1}}}} + \frac{{{C_{{R_{N - 1}}}}}}{{{\Gamma _{N - 1}}{\Gamma _N}}}}}\right)^{-1}
\label{SNDR_1}
\end{align}
where ${d_{N - 1}} = \left( {1 + \kappa _{N - 1}^2} \right)\left( {1 + \kappa _N^2} \right) - 1$ and ${\lambda _N} = 1 + \kappa _N^2$.
The CDF of $\Gamma _{N - 1,N}^F$ can be obtained using \eqref{2-hop-AF-HI} with some variable changes.
Then, the SNDR in \eqref{SNDR_N} can be rewritten as
\begin{align}
\Gamma _{n,...,N}^F = \left({{{d_n} + \frac{{{\lambda _{n,n + 1}}}}{{{\Gamma _n}}} + \frac{{{C_{{R_n}}}}}{{{\Gamma _n}\Gamma _{n + 1,...,N}^F}}}}\right)^{-1},
\label{SNDR_3r}
\end{align}
$\forall n \in [ 1,N - 2]$, where ${\lambda _{n,n + 1}} = {\lambda _{n + 1}} + {C_{{R_n}}}\left( {1 - {\lambda _{n + 1}}} \right)$, ${d_n} = \prod\limits_{i = n}^N {\left( {1 + \kappa _i^2} \right)}  - 1 = {\lambda _n} - 1$, ${\lambda _{N - 1,N}} = {\lambda _N}$
 and
\begin{align}
\Gamma _{n + 1,...,N}^F = \left({{{d_{n + 1}} + \sum\limits_{i = n + 1}^N {{{\lambda _{i + 1}}{\prod\limits_{j = n + 2}^i {{C_{{R_{j - 1}}}}} }}/{{\prod\limits_{j = n + 1}^i {{\Gamma _j}} }}} }}\right)^{-1}
\label{SNDR_23_HI}
\end{align}
Using \eqref{int_f}, we can write the CDF of $\Gamma _{n,...,N}^F$ for $n=1,...,N-2$ as
\begin{align}
{F_{\Gamma _{n,...,N}^F}}(\gamma ) = &{F_{{\Gamma _n}}}\left( {\frac{{{\lambda _{n,n + 1}}\gamma }}{{1 - {d_n}\gamma }}} \right) + \int_0^\infty  {{f_{{\Gamma _n}}}} \left( {y + \frac{{{\lambda _{n,n + 1}}\gamma }}{{1 - {d_n}\gamma }}} \right)
\nonumber\\
\times &
{F_{\Gamma _{n + 1,...,N}^F}}\left( {\frac{{{C_{{R_n}}}\gamma }}{\left({1 - {d_n}\gamma }\right)y}} \right)dy
\label{CDF_3hop_HI_int}
\end{align}
In the first step, we assume $n=N-2$. However, we obtain an integral where there is no closed-form solution. If we use an approximation $ d_n\gamma  \ll 1$ in the denominator of the integral in \eqref{CDF_3hop_HI_int} and follow the same procedure as the proof of \eqref{2-hop-AF-HI}, we obtain the CDF for triple-hop relaying.
Similarly, we can obtain the CDF of $N$-hop relaying with a recursive procedure, where the CDF of ${\Gamma _{n,...,N}^F}$ can be obtained from ${\Gamma _{n+1,...,N}^F}$.
Using the inductive argument, we can obtain the CDF of $N$-hop AF relaying with HI over $\bf H$-fading.
Following the same procedure as the proof of \eqref{dual-hop-HI-exact} and \eqref{2-hop-AF-HI}, we obtain \eqref{N-hop-AF-exact} and \eqref{N-hop-AF-HI}. The details are omitted due to space limitations.

\section*{APPENDIX IV: Proof of Theorem 4}
The asymptotic CDF of MH relaying systems with the imperfection of hardware is provided to evaluate the asymptotic OP and the diversity order.
For the sake of completeness, we don't use the approximation of $ d_n\gamma  \ll 1$.
For $N=2$, using \eqref{SNDR_N}, we can write
 \begin{align}
F_{{\Gamma ^F_{1,2}}}^{{\rm{HI}}}(\gamma)  =& {F_{{\Gamma _1}}}\left( {\frac{{{\lambda _2}\gamma}}{{1 - d_1\gamma}}} \right) + \int_0^\infty  {{f_{{\Gamma _1}}}} \left( {y + \frac{{{\lambda _2}\gamma}}{{1 - d_1\gamma}}} \right)
\nonumber\\ \times&
{F_{{\Gamma _2}}}\left( {\frac{{{C_{{R_1}}}\gamma}}{{\left(1 - d_1\gamma\right)y}}} \right)dt
\label{CDF_HI_dual}
\end{align}
At high SNRs, by applying \cite[Eqs. (1.8.4, 1.8.5)]{Kilbas} to \eqref{cdf_H} and \eqref{pdf_H} and substituting the result into \eqref{CDF_HI_dual} and employing \cite[Eq. (3.194.3)]{Gradshteyn:2007}, the CDF is obtained as in \eqref{CDF_HI_dual2}.
\begin{figure*}
\vspace{-5mm}
 \begin{align}
F_{{\Gamma ^F_{1,2}}}^{{\rm{HI}}}(\gamma) \approx  &
  \sum_{l_1=1}^{\alpha_1}\sum\limits_{i_1 = 1}^{{m_1}}  \frac{{D _{1,i_1}}{\lambda _2^{{\beta _{1,i_1}}}}}{{{{\left( {1 - d_1\gamma} \right)}^{{\beta _{1,i_1}}}}}}{\left(\frac{\gamma}{{\overline \Gamma_1 }}\right)^{{\beta _{1,i_1}}}}+\sum_{l_1=1}^{\alpha_1}\sum_{l_2=1}^{\alpha_2}\sum\limits_{i_1 = 1}^{{m_1}} \sum\limits_{i_2 = 1}^{{m_2}} {{\beta _{1,i_1}}}{{D _{1,i_1}}}{D _{2,i_2}}
 \nonumber\\
 \times & \frac{{\Gamma \left( {1 - {\beta _{2,i_2}}} \right)\Gamma \left( {{\beta _{2,i_2}} - {\beta _{1,i_1}}} \right)}}{{\Gamma \left( {1 - {\beta _{1,i_1}}} \right)}}\frac{{\lambda _2^{{\beta _{1,i_1}} - {\beta _{2,i_2}}}} {C_{{R_1}}^{{\beta _{2,i_2}}}}}{{{{\left( {1 - d_1\gamma} \right)}^{{\beta _{1,i_1}}}}}}\frac{\gamma^{{\beta _{1,i_1}}}}{\left({\overline \Gamma_1 }\right)^{\beta _{1,i_1}}\left({\overline \Gamma_2 }\right)^{\beta _{2,i_2}}}
\label{CDF_HI_dual2}
\end{align}
\hrulefill
\vspace{-5mm}
\end{figure*}
where ${{\beta _{2,i_2}} - {\beta _{1,i_1}}} \notin \mathbb{Z}$, ${\lambda _1}=\left(1+\kappa_1^2 \right)\left(1+\kappa_2^2 \right)$, and ${\lambda _2}=\left(1+\kappa_2^2 \right)$.
\newline
Again, we can use \eqref{CDF_HI_dual2} and derive the CDF of triple-hop relaying as
\begin{align}
F_{{\Gamma ^F_{1,2,3}}}^{{\rm{HI}}}(\gamma)  =& {F_{{\Gamma _1}}}\left( {\frac{{{\lambda _{1,2}}\gamma}}{{1 - d_1\gamma}}} \right) + \int_0^\infty  {{f_{{\Gamma _1}}}} \left( {y + \frac{{{\lambda _{1,2}}\gamma}}{{1 - d_1\gamma}}} \right)
\nonumber\\
 \times &
F_{{\Gamma ^F_{2,3}}}^{{\rm{HI}}}\left( {\frac{{{C_{{R_1}}}\gamma}}{{\left(1 - d_1\gamma\right)y}}} \right)dy
\label{CDF_HI_dual3}
\end{align}
where $F_{{\Gamma ^F_{2,3}}}^{{\rm{HI}}}\left(\gamma \right)$ is the same as \eqref{CDF_HI_dual2} with parameter change of indices from 1 and 2, to 2 and 3, respectively.
Using \cite[Eq. (3.197.1)]{Gradshteyn:2007}, we can solve the integral in \eqref{CDF_HI_dual3} and obtain
 \begin{align}
{F_{{\Gamma ^F_{1,2,3}}}^{\rm HI,\infty}}(\gamma) \approx  \sum_{l_1=1}^{\alpha_1}\sum\limits_{{i_1} = 1}^{{m_1}} {\frac{{{{D_{1,{i_1}}}}{A_3}}}{{{{\left( {1 - {d_1}\gamma } \right)}^{{\beta _{1,{i_1}}}}}}}} {\left( {\frac{\gamma }{{{{\overline \Gamma }_1}}}} \right)^{{{\beta _{1,i_1}}}}}
\end{align}
where $A_3$ is defined in \eqref{CDF_HI_dual4} with ${{\beta _{3,i_3}} - {\beta _{2,i_2}}} \notin \mathbb{Z}$.
\begin{figure*}
\vspace{-1mm}
\begin{align}
{A_3} \triangleq & \hspace{2mm}
  \lambda _{2}^{{\beta _{1,{i_1}}}}+\sum_{l_2=1}^{\alpha_2}
{\sum\limits_{{i_2} = 1}^{{m_2}} {{ {} }} }{\frac{ D _{2,{i_2}} \lambda _{3}^{{\beta _{2,{i_2}}}}}{{{\beta _{2,{i_2}}} - {\beta _{1,{i_{1}}}}}}} {\frac{{{{\left( {{\lambda _{1,2}}} \right)}^{{\beta _{1,{i_{1}}}}}}}}{{d_{2}^{{\beta _{2,{i_2}}}}}\left({\overline \Gamma_2 }\right)^{\beta _{2,i_2}}}}
{{}_2{F_1}\left( {{\beta _{2,{i_2}}},1;{\beta _{2,{i_2}}} - {\beta _{1,{i_{1}}}} + 1,1 + \frac{{{\lambda _{1,2}}}}{{{d_{2}}{C_{{R_{1}}}}}}} \right)}
 \nonumber\\
 + &\sum_{l_2=1}^{\alpha_2}\sum_{l_3=1}^{\alpha_3}\sum\limits_{{i_2} = 1}^{{m_2}} {\sum\limits_{{i_3} = 1}^{{m_3}}}{}\frac{{{\beta _{2,{i_2}}}}{{D _{2,{i_2}}}}D _{3,{i_3}}C_{{R_{2}}}^{{\beta _{3,{i_3}}}}}{\lambda _3^{{\beta _{3,{i_3}}}-{\beta _{2,{i_{2}}}} }}
\frac{{\Gamma \left( {1 - {\beta _{3,{i_3}}}} \right)\Gamma \left( {{\beta _{3,{i_3}}} - {\beta _{2,{i_{2}}}}} \right)}}{{\Gamma \left( {1 - {\beta _{2,{i_2}}}} \right)} }\frac{{\left( { - 1} \right)^{ { - {{\beta _{2,{i_2}}}} }}}}{{d_{2}^{{\beta _{2,{i_2}}}}}}
 \nonumber\\
\times &
{\frac{{{{\left( {{\lambda _{1,2}}} \right)}^{{\beta _{1,{i_{1}}}}}}}\left({{{\beta _{2,{i_2}}} - {\beta _{1,{i_{1}}}}}}\right)}{\left({\overline \Gamma_2 }\right)^{\beta _{2,i_2}}{\left({\overline \Gamma_3}\right)^{\beta _{3,i_3}}}}}{{}_2{F_1}\left( {{\beta _{2,{i_2}}},1;{\beta _{2,{i_2}}} - {\beta _{1,{i_{ 1}}}} + 1,1 + \frac{{{\lambda _{1,2}}}}{{{d_{2}}{C_{{R_{1}}}}}}} \right)}
\label{CDF_HI_dual4}
\end{align}
 \hrulefill
\vspace{-5mm}
\end{figure*}
Following an inductive argument, we can derive the asymptotic OP for $N$-hop relaying. The closed-form expression is given in \eqref{CDF_HI_N_hsnr}.

%

 \bibliographystyle{IEEEtran}
\bibliography{IEEEabrv,reference}
\vspace{-1cm}
 \begin{IEEEbiography}[{\includegraphics[width=1in,height=1.25in,clip,keepaspectratio]{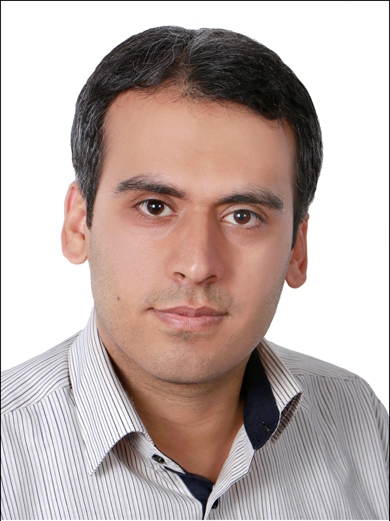}}]
{Ehsan Soleimani-Nasab} received the B.Sc. degree in electrical engineering from the Iran University of Science and Technology, Tehran, Iran, in 2006, and the M.Sc. and Ph.D. degrees in ommunication systems from the K. N. Toosi University of Technology, Tehran, in 2009 and 2013, respectively. From April 2012 to October 2012,
he was a Visiting Researcher with the Department of Signals and Systems, Chalmers University of Technology, Gothenburg, Sweden. From June 2014 to August 2014, he was worked as a Research Associate with the Department of Electrical and Electronics Engineering, Özyegin University, Istanbul, Turkey. Since September 2014, he has been with the Graduate University of Advanced Technology, Kerman, where he is currently an Associate Professor. From October 2022 to September 2023, he was a Guest Researcher at
the Department of Electrical and Electronics Engineering, Koc University, Istanbul, Turkey. He is the author or coauthor of around 60 journal and conference publications. His research interests include optical wireless communications, radio wireless communications, and signal processing in communications. He has served on the technical program committees for various IEEE conferences. He is an active Reviewer for various the IEEE TRANSACTIONS and other journals.
\end{IEEEbiography}
\vspace{-5mm}
\begin{IEEEbiography}[{\includegraphics[width=1in,height=1.25in,clip,keepaspectratio]{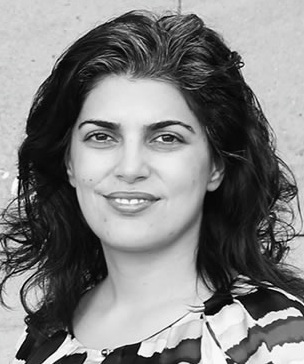}}]
{Sinem Coleri} is a Professor and the Chair of the Department of Electrical and Electronics Engineering at Koc University. She is also the founding director of Wireless Networks Laboratory (WNL) and director of Ford Otosan Automotive Technologies Laboratory. Sinem Coleri received the BS degree in electrical and electronics engineering from Bilkent University in 2000, the M.S. and Ph.D. degrees in electrical engineering and computer sciences from University of California Berkeley in 2002 and 2005. She worked as a research scientist in Wireless Sensor Networks Berkeley Lab under sponsorship of Pirelli and Telecom Italia from 2006 to 2009. Since September 2009, she has been a faculty member in the department of Electrical and Electronics Engineering at Koc University. Her research interests are in 6G wireless communications and networking, machine learning for wireless networks, machine-to-machine communications, wireless networked control systems and vehicular networks. She has received numerous awards and recognitions, including N2Women: Stars in Computer Networking and Communications, TUBITAK (The Scientific and Technological Research Council of Turkey) Incentive Award and IEEE Vehicular Technology Society Neal Shepherd Memorial Best Propagation Paper Award. Dr. Coleri has been Interim Editor-in-Chief of IEEE Open Journal of the Communications Society since 2023, Executive Editor of IEEE Communications Letters since 2023, Editor-at-Large of IEEE Transactions on Communications since 2023, Senior Editor of IEEE Access since 2022 and Editor of IEEE Transactions on Machine Learning in Communications and Networking since 2022. Dr. Coleri is an IEEE Fellow and IEEE ComSoc Distinguished Lecturer.
\end{IEEEbiography}
\end{document}